\documentstyle[preprint,tighten,eqsecnum,epsf,aps]{revtex}

\def\beq{\begin{equation}}   \def\eeq{\end{equation}}

\begin{document}
\draft
\title{Parton-Hadron duality in QCD sum rules: quantum mechanical examples}
\author{
B. Blok\thanks{E-mail: PHR34BB@vmsa.technion.ac.il} and
M. Lublinsky\thanks{E-mail: mal@techunix.technion.ac.il}}
\address{Department of Physics, Technion -- Israel Institute of 
Technology, Haifa 32000, Israel}
\maketitle

\thispagestyle{empty}

\begin{abstract}
Motivated by recent work on three-point QCD sum rules
in heavy quark physics, we use the simple quantum mechanical models
to study the basic issue of duality in three-point sum rules. We show
that while in all of these models the duality in two-point sum rules works 
fine, the duality in three-point sum rules may be 100$\%$- violated,
leading to completely unreliable predictions for the matrix elements in 
question. The implications for  three-point QCD sum rules are discussed.
A new estimate for the parameter $\lambda_1$ of HQET is given.
\end{abstract}


\setcounter{page}{1}
\section{Introduction} 

Recently there has been growing interest in the issue of parton-hadron 
duality in QCD \cite{Shifman,BM,Chib,Blok}. 
This is due to both theoretical and 
experimental progress. On the one hand, we witness  further 
improvements in 
theoretical methods of  investigation of  low energy properties of 
hadrons, such as the QCD sum rules and  Heavy Quark Effective 
Theory.
 These methods enable one to calculate the hadronic 
properties directly from QCD, in a model independent way. 
 These methods  heavily rely 
on   the validity of the parton hadron duality.

In addition to theoretical advances, one can now compare the predictions 
of these methods with the growing amount of  the new experimental data. 
This leads naturally to the renewed 
interest in the fundamentals of these methods, i.e.,  in the issue of 
how reliable they are.

 One especially important application of the idea of the parton-hadron 
duality is the QCD sum rules \cite{SVZ} (see also Ref. \cite{RRY} for  
review).
 For the past 20 years  QCD sum rules have been widely
and successfully  used to predict 
masses (the so called two-point QCD sum rules) and the coupling constants
(three-point QCD sum rules) of different hadrons and their decays.

The basic procedure in  QCD sum rules is the following one: 
one calculates the 
physical quantity - the polarisation operator of a certain number of 
currents in two ways.
 First, we calculate the polarisation operator
 in terms of quarks and gluons, using asymptotic 
freedom. Then, we  calculate the same polarisation operator 
in terms of hadrons, using  dispersion relations. One then equates 
the results from the parton model (the so called theoretical part of the
QCD sum rule) with the sum over the hadron states (the so called 
phenomenological part of the sum rule).
 Usually, we are interested only in the properties of the lowest lying 
resonance.
The contribution of the higher  resonances
created by the given currents is taken into account using  
the so called continuum model \cite{Ioffe}. In other words, we approximate 
the hadron spectral density
(i.e., the imaginary part of the polarisation operator) 
 by some smeared function. The standard approach is to
assume that the corresponding smeared function is well 
approximated by the spectral density of the theoretical part of the sum 
rule. The assumed validity of this approach underlies  all
practical QCD 
sum rule calculations. It is exactly the issue of whether the parton hadron 
duality holds.  

The issue of the parton hadron duality and the
closely related issue of the  right model for the 
continuum are not new. The subject was discussed in detail in the early paper 
by Shifman, Vainshtein and Zakharov
\cite{SVZ}, following the classical discussion by Poggio, Quinn and 
Weinberg \cite{PQW} of the parton hadron duality in the case of the $e^+-e^-$ 
annihilation.   
 
Unfortunately, one cannot rigorously check the parton hadron duality
directly 
in QCD. Although many  arguments support  duality,
one cannot tackle the issue without the explicit theory of confinement. 
Consequently, one has to check the hadron parton duality (or, rather, its 
analogues) in simpler models. The existence (nonexistence) 
of  duality in these models is a strong argument for  (against) the 
parton hadron duality in real QCD.
 
While we are still unable to study  the issue of duality directly in QCD, one 
can learn a lot by  studying various exactly solvable 
models, the simplest of which are the quantum mechanical potential models.
These models have been used to  gain insight into the issue of the 
parton hadron duality
in the case of  two-point QCD sum rules \cite{NSVZ,Durand,Bert}.   

 The goal of the present paper is to study 
the analogue of the parton-hadron  duality for three-point sum rules 
in the quantum-mechanical potential models. 
The issue has become especially relevant recently,
due to the extensive use of  three-point QCD sum rules for the determination 
of different parameters of  Heavy Quark Physics (see e.g. the review  
\cite{N}). In particular,
it was 
found that different QCD sum rules lead to contradictory values of 
several fundamental parameters of 
Heavy Quark Effective Theory (HQET), like 
$\langle B |\vec{D}^2/(2m_B)| B\rangle$
\cite{Neubert,BB};  these values, in turn, differ from the ones 
predicted on the basis of the analysis of the experimental data
\cite{KLW,Chernyak}.
This requires us to go back and check once again the 
basic assumptions behind the QCD sum rule method.
 
Our present study  confirms the old results \cite{NSVZ,Durand,Bert} 
that state that the
duality works excellently for the two-point sum rules. However, for the same 
models where the
two-point duality works excellently, we shall see that we may encounter 
serious 
problems in the study of the three-point sum rules. Namely, we shall 
see that not only local duality, but also generalised duality (in the sense
defined in Ref. \cite{BS}) are violated in the  situation when 
the two-point sum rules work excellently.

\noindent 
We shall consider in this note three basic models:

\noindent
A) The harmonic oscillator.
This is the potential model described by the potential
\beq
V(r)=\omega r^2/2.
\label{harmonic}
\eeq
B) The linear oscillator. This model is the basis of the realistic
potential models;
\beq
V(r)=\alpha r.
\label{linear}
\eeq
C) The last model to consider is the linear potential perturbed 
by the coulombic interaction that imitates the effects of the $\alpha_s$
corrections in the potential models:
\beq
V(r)=\alpha r -\beta /r.
\label{coulombic}
\eeq
 The three-point sum rules for the harmonic oscillator 
 were already studied in Refs. \cite{BS,BSU}. (The numerical mistake
 made in Ref. \cite{BS} in the three-point sum rule for oscillator
(but not, of course, for the QCD sum rules for the slope of Isgur-Wise 
function)
was improved in Ref. \cite{BSU}).
In Ref. \cite{BS}, it was shown that the duality may hold 
for the harmonic oscillator in generalised sense only: a one-dimensional
 integral of the phenomenological
spectral density is dual to the corresponding integral of the theoretical 
part of the sum rule.  However, later investigation
\cite{BSU} showed that even 
this duality does not take place, and the three-point sum rules for 
harmonic oscillator do not reproduce the true values. Here, we shall see
that the harmonic oscillator case is not an exception but the general 
situation. The duality in three-point sum rules breaks down because
the coupling signs become alternating. In all three models in question,
the duality breaks for the sum rules determining ground state matrix 
elements of the following operators:
\beq
O_1\sim r^2,\ \ \ \ \ \ \ \ \ \ O_3\sim r.
\label{operators}
\eeq
(Note the close analogy between the operator $O_1$ and the operator
that determines the fundamental parameter of the HQET - the slope of 
the Isgur-Wise function).  

As a result of the duality breaking, the standard continuum model
\cite{Ioffe} does not describe the true spectral density for the first
several resonances. So, the sum rules give answers that 
differ from the right ones (here, in quantum mechanics, we know, of course,
the exact values) by $30-50\%$.
 On the other hand, in the case of the kinetic energy operator 
$O_2\sim -\partial^2$, all nondiagonal transitions give positive 
contributions. The sum rules do work and the 
duality is not broken. Nevertheless, 
there is a big continuum contribution leading to large uncertainties in the 
predictions.   

This paper is organised as  follows. In section II,  we review 
the notion of duality and discuss the duality in the 
two-point sum rules for the
potential models mentioned in the Introduction. In section III,  
we study the duality in the three-point 
quantum mechanical sum rules. The duality fails for the operators
$\hat{O}_{1}, \hat{O}_{3}$, and we trace the origin
of its failure. On the other hand, for the operator $\hat{O}_{2}$ the duality
holds, and we investigate the corresponding sum rules.
 In section IV, we give our
conclusions and discuss  possible implications of our results for QCD.
The details of the exact solution of the quantum mechanical models at hand 
are  given in Appendix A, while the details of the calculations of the 
theoretical part of the sum rules are given in Appendix B.

\section{Duality and two-point sum rules}

Let us recall now in more detail 
what do we mean by quark hadron duality in the case of the QCD sum rules.
Consider the function $f(q^2)$
at some tensor structure in the polarisation operator 
of two currents in the sum rule.
One calculates $f$  by means of the Operator Product Expansion 
in the Euclidean domain of momenta $q^2\le 0$:
\beq
f(q^2)\sim a_0 +a_1/q^4+\cdots .
\label{f}
\eeq
Here the coefficient $a_0$ corresponds to the perturbation theory 
and the coefficients $a_i, i\ge 1$ correspond to the matrix elements of 
the relevant operators over the QCD vacuum. (Possibly, the coefficients
$a_0, a_i$ depend logarithmically on $q^2$).
The function $f$ is a smooth function of its argument  $q^2$. 
Its imaginary part, denoted  $\sigma_{\bf p} (s)$, is
 a smooth function as well.
Another way to calculate  $f$ is to express it  in terms of hadron 
properties by means of the dispersion relation:
\beq
f(q^2)=\int \frac{{\rm Im} f(s)}{(s-q^2)}.
\label{disp}
\eeq
The imaginary part $\sigma_h(s)\equiv {\rm Im}f(s)$ involved 
equals the sum of the delta functions over  hadronic resonances
contributing into the tensor structure in question:
\beq 
\sigma_h (s)=\sum_n \beta_n^2\delta (s-m^2_n).
\label{delta}
\eeq
Now, the parton-hadron duality assumes that, starting at some threshold $s_1$,
the integrals of the hadronic spectral density $\sigma_h(s)$ and the 
partonic one, $\sigma_p (s)$, with exponential weights coincide:
\beq
\int_{s_1}^\infty \sigma_{\bf p}(s)e^{-s/M^2}\sim
\int_{s_1}^\infty \sigma_h(s)e^{-s/M^2}.
\label{duality}
\eeq

The minimal $s_1$ satisfying this equation is  is called the
duality threshold. For  QCD sum rules to work, $s_1$ must lie between 
the masses of the ground state 
 and the first excited state contributing to $f$.
In other words, the hadron spectral density, averaged over some 
interval of $s$, must be approximately equal to the spectral density 
calculated in the parton model. Of course, this is  true  for  
the exact spectral densities. However,
our $\sigma_p (s)$  is only a part of the exact spectral 
density, namely the part  corresponding to the first several 
terms in the OPE, analytically continued to Minkowsky space.

 Once we know what the duality  means in QCD, let us consider 
its quantum mechanical analogue.
The quantum mechanical analogue of the polarisation operator of two
currents in QCD is  the time dependent Green function:
\beq
S_0(0,T)=\sum_n\vert \psi_n (0)\vert^2e^{-E_nT}.
\label{analogue}
\eeq
 Here $n$ runs  over the S-states only. This function describes the 
propagation from the point (0,0) to the point (0,T) in the Euclidean 
time. One can calculate the function $S_0$  in two ways. First, we can 
calculate it as a power series in $T$ for short times. The leading
term in such a series is an analogue of the parton model contribution
in QCD; the higher terms imitate the matrix elements of operators in 
the operator product expansion. Second, one can use the explicit
formula (\ref{analogue}) and calculate $S_0$ as a sum over the hadron 
states. In practice, we are interested in the properties of the 
ground state.
So we represent 
\beq
S_0 (0,T)=\vert \psi_0 (0)\vert^2 e^{-E_0T}+\int_{E_c}^{\infty}dE\sigma_h(E)
e^{-ET},
\label{spect}
\eeq
where $\sigma_h (E)$ is the exact spectral density,
\beq
\sigma_h (E)=\sum_{n=1} \delta (E-E_n)\vert \psi_n (0)\vert^2.
\label{sigma}
\eeq
Note that the leading term in the perturbation expansion of the $S_0(T)$
can also be represented  as the integral of the spectral 
density $\sigma_{\bf p} (E)$ times $e^{-ET}$. 
Then the quantum mechanical duality
means that, after average over some energy interval,
\beq
\sigma_h (E)\sim \sigma_{\bf p} (E),\,\,\,\,\ \  E\ge E_c.  
 \label{average}
\eeq
 Equivalently, the integrals of the spectral functions times $e^{-ET}$
are approximately equal as functions of $T$.
 The integral of the
 exact spectral density in the r.h.s. can be also calculated
as the difference between the exact Green function $S_0 (T)$ and the 
known exact expression for the ground state. Define 
\beq
C_0 (  T)= \frac{S_0(0,T)-\vert \psi_0 (0)\vert^2e^{-E_0T}}
{\vert \psi_0 (0)\vert^2e^{-E_0T}},
\label{exact}
\eeq
and
\beq
C_{\rm p}(  T)=\frac{\int_{E_c}^\infty dE\sigma_{\bf p}(E)e^{-ET}}
{\vert \psi_0 (0)\vert^2e^{-E_0T}}.
\label{exact1}
\eeq
Below the functions $C_0(T)$ and $C_{\rm p}(T)$ are called ``continuum''
functions. If the parton hadron duality holds, these two functions must 
approximate one another for sufficiently small $T$
(corresponding to sufficiently large  $E$). Once duality is established, 
one can write the sum rule to determine the ground state parameters. In order
to obtain the sum rule we simply rewrite equation (\ref{spect}) differently:
\beq
\vert \psi_0 (0)\vert^2e^{-E_0T}=S_0(0,T)-
\int_{E_c}^\infty dE\sigma_{\bf p}(E)e^{-ET}.
\label{specnew}
\eeq
Here, for $S_0$ we use the power expansion, which can be obtained in 
Perturbation Theory. By fitting the r.h.s. of equation (\ref{specnew}) to
exponent one is able to determine both ground state energy $E_0$ and its
residue $\vert \psi_0 (0)\vert^2$. The ground state energy can be easily
obtained by first taking logarithm of (\ref{spect})  and then differentiating
with respect to time $T$. Since our method is an approximate one, we have to
introduce a notion of fiducial region. By fiducial domain we mean a window
in $T$ where two following conditions hold simultaneously. The first one is
a control over the power expansion of $S_0$.
Usually one demands for the last kept
term in the expansion to saturate less than 30\% of the whole expression.
This way the upper edge of the window is determined. The second condition
is a ground state dominance. This requirement is needed in order to suppress
the relative contribution of the exited states. This condition determines
a lowest edge of the window. Practically  the contribution of all exited
states (the integral term in (\ref{specnew})) is required 
to be less than 30\%. The fiducial domain corresponds to the region where
two asymptotics (small $T$ and large $T$) matches. The important fact is that 
the sum rule (\ref{specnew}) are essentially threshold dependent. Usually
we do not know the value of the threshold parameter $E_c$ except the general
point that it should lie somewhat below the energy of the first exited state
(which we do not know too). However, the standard philosophy of sum rules is
to seek for a region (in $E_c$) where $E_c$ dependence is small. The sum rule
is then called stable. The variation of the result with $E_c$ produces an 
error, which is unavoidable in the sum rule  method. Two-point
sum rules discussed below happen to be very stable with respect to the
threshold variation. We depict the sum rules with only one optimal value
for the threshold parameters, which we define  by the best fit to exact known 
results. In all models, the optimal values appear to be very close to the 
guesses typical in practice, e.g., the midpoint between the lowest observed
states.

Let us illustrate the quantum mechanical duality for two-point 
sum rules for the A, B, C models discussed in  Introduction
\cite{NSVZ,Durand,Bert,BS,BSU}. Although similar discussions are already 
present in the literature, we shall also consider 
these models for the sake of completeness and  
as  simple illustrations of more complicated cases of Section III.
 
Consider the harmonic oscillator first. We use the
 dimensionless units $\omega=1$ and $2m=1$.
The left, theoretical part of the sum rule can be represented as a 
perturbation series:

\begin{equation}
S_0^{\rm har}(0,T)=\left(\frac{1}{4\pi T}\right)^{3/2} 
\left(1-\frac{1}{4} T^2
+\frac{19}{480}  T^4-\frac{691}{120960}  T^6 +\cdots\right).
\label{leftA}
\end{equation}
The corresponding spectral function is

\beq
\sigma_{\bf p} (E)=\frac{1}{4\pi^2}\sqrt E.
\label{partA}
\eeq

Consider now the right hand side of the sum rule. The spectral density
$\sigma_h(E)$ can be represented as a sum of delta functions.
For sufficiently high energies  the summation can be approximately 
substituted by the
integration \cite{NSVZ,BS}. Using $E_n=3/2 +2n $ and 
$|\psi_n(0)|^2=\left(\frac{1}{2\pi}\right)^{3/2}\frac{(2n+1)!!}{2^nn!}$
, one  obtains

\begin{eqnarray}
\sigma_h^{\rm har}(E)\sim & &\sum_n 
\left(\displaystyle{\frac{2}{\pi}}\right)^{1/2} 
\displaystyle{ \frac{E_n^{1/2} }{(2\pi)^{3/2}}}
\delta (E-E_n)\approx \nonumber \\ \nonumber \\
& &\int \left(\frac{2}{\pi}\right)^{1/2} 
E_n^{1/2} \left(\frac{1 }{2\pi}\right)^{3/2} 
\delta (E-E_n)\frac{dE_n}{2 }=
\frac{1}{4\pi^2} \sqrt E. 
\label{sum}
\end{eqnarray}
The latter is just the theoretical spectral density  
(\ref{partA}). We see that for sufficiently high energies the duality  
holds indeed.
 
We can make more detailed estimates, relevant for the sum rules.
 Define the continuum functions $C_0^h(T)$ and $C_{\rm p}^h(T)$ 
(\ref{exact}),(\ref{exact1}).
Here we use the exact solutions for the harmonic oscillator:
$E_0^{\rm har}=\frac{3}{2}$;  
$\vert \psi_0^{\rm har} (0)\vert^2=\left(\frac{1 }{2\pi}\right)^{3/2}.$
These two functions approximate each other very  well for $T\le 1.7$, 
(see Fig.\ \ref{S0},
where the two functions are depicted for the optimum value of $E_c=2.6$).
Once duality is established, one can write the sum rule 
to determine the ground state parameters. The fiducial region 
in this sum rule must, of course, be inside the
region where the duality holds.
Straightforward calculations (see
Refs. \cite{NSVZ} for details) show that this is true indeed,
and the resulting values coincide with the exact ones
 very  accurately.

Consider now the case of the linear oscillator. We use units 
where $\alpha=1$.
The l.h.s. of the sum
rule is given by the asymptotic expansion of the propagator in Perturbation
Theory. The corresponding Green function was obtained in Ref.\ \cite{WDD}
 (see also  Appendix B):
\begin{eqnarray}
  \label{sl0}
  S^L_0(0,T)=\frac{1}{(4\pi T)^{3/2}}
\left(1-\frac{\sqrt{\pi}}{2}T^{3/2}+
\frac{5}{12} T^3 + \cdots \right),
\end{eqnarray}
whereas the corresponding parton spectral function 
is given by eq. (\ref{partA}).
Here and below, the index $L$ denotes the linear oscillator problem.

Pass now to the r.h.s. of the sum rule. Proceeding in the
same spirit as for the harmonic oscillator, the local duality for
high energies can be  established. We use some exact results on the 
problem,  collected in Appendix A. The square of the wave-function
$|\psi^L_n(0)|^2$ equals $1/(4\pi)$. 
The large $n$ asymptotic behaviour of 
the energy levels  $E_n^L=(3/2 \pi n)^{2/3}$. 
The level density $\sigma^L_h(n)$ is 
\begin{eqnarray}
  \label{speclin}
\sigma^L_h(n) \sim \frac{1}{4\pi}\sum_k \delta(k-n) \approx  
\frac{1}{4\pi}\int \delta(k-n) dn=\frac{1}{4\pi}.
\end{eqnarray}
The energy density is 
$\sigma^L_h(E)=\sigma^L_h(n)\frac{\partial n}{\partial E}$, which
 coincides with the ``bare'' spectral function (\ref{partA}).

The duality for sufficiently high energies established, we turn to 
  the sum rule.  The appropriate 
``continuum'' functions $C^L_0( T)$ and 
$C_{\bf p}(T)$
of (\ref{exact}) and (\ref{exact1}) are 
depicted in Fig.\ \ref{SL0} for the optimal energy threshold 
$E_c=3.4 $.  The fit is perfect and we can plot the sum rule
for the ground state energy (Fig.\ \ref{fig2a}). One obtains 
 $E^L_0\approx (2.35\pm 0.05)$, while the exact value is
$E^L_{0\rm exact}=2.338$. 
We see  the sum rules really do work!

Our last example of two-point functions, is C model with its
 ``Linear + Coulomb'' potential. 
The model and its numeric  solution are 
described  in Appendix A, while the perturbative expansion of the propagator
is presented in Appendix B. Below,  numerical solutions of the model
will be referred to as exact ones.

\begin{eqnarray}
  \label{slc0}
S_0^{\rm lc} (0,T)=\frac{1}{(4\pi T)^{3/2}}\left(1+b_0\sqrt{\pi} T^{1/2}+
b_0^2 \frac{\pi^2}{6}T- b_0\frac{3}{2}T^2 -\frac{\pi^2}{2}T^{3/2}+
\frac{5}{12}T^3+\cdots \right).
\end{eqnarray}

The parameter $b_0=0.57$ is defined in Appendix A.  In QCD, parton spectral 
density acquires  corrections due to $\alpha_s$  terms in the OPE. 
In C  model, the Coulomb interaction imitates the role of these
$\alpha_s$ terms. The parton spectral function (\ref{partA}) is modified:

\begin{equation}
  \label{speclc0}
  \sigma_{\bf p}^{\rm lc}(E)= \frac{1}{4\pi^2} E^{1/2}+
\frac{b_0}{8\pi}+\frac{b_0^2}{48} E^{-1/2}.
\end{equation}

Unfortunately, contrary to  the previous  examples we do not know
exact solutions of the problem. However, it is natural to believe that
the large $n$ asymptotic behaviour of the wave-function 
and the energy spectrum can be obtained in WKB based methods.
We believe that this way one can confirm that the duality holds in fact.  

Following the procedure described above, the continuum functions 
$C^{\rm lc}_0 (T)$ and $C_{\bf p}^{\rm lc}(T)$  (\ref{exact})
and (\ref{exact1}) 
were computed  and plotted for the optimal 
threshold $E_c=2.9$ (Fig.\ \ref{SLC0}).
Within the window ($0.2\le T\le 0.9$), the duality is valid and we can
study the sum rules. Fig.\ \ref{fig3a} presents the sum rule for the ground
state energy. 
The sum rule result $E^{\rm lc}_0=1.90 \pm 0.05$ matches well the
exact one $E^{\rm lc}_{0\rm exact}=1.83$.

\section{Duality in three-point sum rules}

The goal of the present section is to study the issue of the duality 
in the  three-point sum rules. While we have seen in the previous chapter
that the duality holds for the two-point sum rules,
 (in quantum mechanics at least),
the situation for the three-point sum rules is clearly more complicated.
Indeed, let us recall the general procedure of the analysis of three-point
sum rules in QCD \cite{NR}. One considers the
function $f(q_1^2,q_2^2)$ at the appropriate tensor structure of the
 polarisation operator of three 
currents. One can calculate this function in two ways:
using  Operator Product Expansion (the theoretical part of the sum rule)
and saturating by resonances (the phenomenological part of the sum rule).
For simplicity we shall restrict ourselves here by transitions between the 
same hadron under the 
action of some current.
Then, after the Borel transformation in variables $q_1^2, q_2^2$ the 
phenomenological part of the sum rule can be represented as
\beq
f(M_1^2,M^2_2)=g\beta_0^2 e^{-m^2(1/M_1^2+1/M_2^2)}+\int ds_1ds_2
\sigma_h (s_1,s_2)e^{-s_1/M_1^2-s_2/M_2^2}.
\label{three}
\eeq
Here $g$ is the relevant coupling constant, $\beta_0^2$ is the square of
the residue of the lowest lying resonance created by the current, 
$m$ is its mass  and $M_1^2, M_2^2$ are the relevant Borel parameters. 
Local duality means that the latter integral, taken over 
some part of the $(s_1,s_2)$ plane is well approximated by the 
corresponding integral of the imaginary part $\sigma_{\bf p}(s_1,s_2)$
of the theoretical part 
of the sum rule (calculated using Wilson OPE). 
Even if the local duality does not hold, for the three-point sum rules it is
possible to have the generalised duality. In fact, if there are sign 
alternating
transition, it was argued in Ref. \cite{BS} that it may be senseless  
to speak about local duality.
The parton model density is likely to be concentrated in the narrow 
area around the diagonal of the ($s_1, s_2$) plane, while the hadron density
is spread over the whole plane. It was shown in Ref. \cite{BS} that in this
case  only  ``generalised'' duality makes sense:
the partonic spectral density, integrated in the direction orthogonal 
to the diagonal is approximately equal to the hadron spectral density,
integrated in the same direction:
\beq
f(M^2)=g\beta^2_0 e^{-m^2_0(1/M^2)}+\int^{\infty}_{s_0} ds
\sigma_h (s)e^{-s/M^2}.
\label{three1}
\eeq
Here $\sigma_h(s)$
is given by the integral
\beq
\sigma_h (s)=\int dA \sigma_h (s_1,s_2),
\label{four}
\eeq
$A=(s_1-s_2)/2$. The parameter
$s_0$ is the continuum threshold (see Fig.\ \ref{spec}). The Borel parameters 
are $M^2=M_1^2=M_2^2$ and we stick to the symmetric point.
 We can define the 
parton spectral density  exactly in the same way:
\beq
f(M^2)=\int \sigma_{\bf p}(s)e^{-s/M^2}ds
\label{five}
\eeq
 Only for such sum rule there is a  hope that the 
duality (defined in this generalised sense) is not be violated.

Consider now the quantum mechanical analogue of the sum rule (\ref{three1}).
The analogue of the polarisation operator of three currents in quantum 
mechanics is the function \cite{BS}:

\begin{eqnarray}
  \label{kok}
S_i(\tau_1,\tau_2)=\int &d^3&r K(0,\tau_1+\tau_2,r,\tau_1)\hat{O}_i(r)
K(r,\tau_1,0,0)= \nonumber \\
\int &d^3&r \sum_l e^{-E_l\tau_2}\psi_l(0)\psi_l^*(r)\hat{O}_i(r)
 \sum_n e^{-E_n\tau_1}\psi_n(r)\psi_n^*(0).
\end{eqnarray}
Here $K(r,\tau_1,0,0)$ is the amplitude of the quark propagation from the
point $(0,0)$ to the point $(r,\tau_1)$ in the Euclidean time. At the point
$(r,\tau_1)$ the operator $\hat{O}_i$ is inserted.

The vacuum expectation value (VEV) of the operator $\hat{O}_i$ is defined as
\begin{equation}
  \label{32}
\langle 0|\hat{O}_i|0\rangle = \int d^3r \psi_0^*(r)\hat{O}_i(r)\psi_0(r).
\end{equation}
The corresponding sum rule can be written as:
\begin{equation}
  \label{sumtri}
S_i(\tau_1,\tau_2)=|\psi_0(0)|^2 \langle 0|\hat{O}_i|0\rangle 
e^{-E_0(\tau_1+\tau_2)} +
\int_{s_0}^\infty ds_1 ds_2\sigma_h (s_1,s_2) e^{-s_1\tau_1-s_2\tau_2}.
\end{equation}
Here, $\sigma_h (s_1,s_2)$ is the exact spectral density. By 
$\sigma_{\bf p} (s_1,s_2)$ 
we denote the theoretical spectral function obtained, 
as for the two-point functions, from the $T$-expansion of $S_i$.
From  above it is clear that we must study the symmetric sum rules,
i.e. $\tau_1=\tau_2 =T/2$
 (see Ref. \cite{BS} for details).
We  need to compare the $\sigma_{\bf p} (s)$ integrated 
with the weight $ e^{-sT}$, with the corresponding integral of $\sigma_h (s)$.
Here $s=(s_1+s_2)/2$, and $\sigma_{h,p} (s)$ are the spectral densities 
obtained
from the spectral densities $\sigma_{h,p} (s_1, s_2)$ after the 
integration over
the variable $A=(s_1-s_2)/2$ in the same way as in eq. (\ref{four}).

Let us now consider the sum rules and duality for three models 
considered above and for the operators:
\begin{eqnarray*}
  \hat{O}_1(r)=r^2/6; \ \hat{O}_2(r)=-\partial^2; \ \ \hat{O}_3(r)=r.
\end{eqnarray*}

\subsection{Harmonic oscillator}

Since the operator $\hat{O}_3$ has no analogue in QCD, the sum rules only
for two operators $\hat{O}_1$ and 
$\hat{O}_2$ will be discussed. These sum rules were already investigated 
in Refs. \cite{BS,BSU}. For the sake of completeness their 
analysis is included and extended. 
\begin{eqnarray}
  S_1^{\rm har}(T/2,T/2)&=&\frac{1}{32\pi^{3/2}}\frac{1}{T^{1/2}}
\left(1-\frac{1}{3} T^2 +\frac{44}{640} T^4 -\frac{692}{60480} T^6+
\cdots\right);
\nonumber \\ \label{s1h} \\
 S_2^{\rm har}(T/2,T/2)&=&\frac{3/2}{(4\pi)^{3/2}}\frac{1}{T^{5/2}}
\left(1-\frac{1}{6} T^2 +\frac{5}{288} T^4 +\cdots\right). \nonumber
\end{eqnarray}
The corresponding parton spectral density is
\begin{eqnarray}
\sigma^{\rm har}_{\rm p1}(E)&=&\frac{1}{32\pi^2}E^{-1/2};  \nonumber \\ 
\label{spech}
\sigma^{\rm har}_{\rm p2}(E)&=&\frac{3/2}{(4\pi)^{3/2}}
\left(\frac{4}{3\sqrt{\pi}}E^{3/2}-
\frac{1}{6\sqrt{\pi}}E^{-1/2}\right).
\end{eqnarray}
In order to check duality let us, as for the two-point functions, eqs.
(\ref{exact}) and (\ref{exact1}), define the appropriate continuum 
functions: 
\begin{eqnarray}
  \label{cont}
  C_i(T)&=& \frac{S_i(T/2,T/2)-
\vert \psi_0 (0)\vert^2e^{-E_0T}\langle0|\hat{O_i}|0\rangle}
{\vert \psi_0 (0)\vert^2e^{-E_0T}\langle0|\hat{O_i}|0\rangle}; \nonumber \\
\\
C_{\rm pi}(T)&=&\frac{\int_{E_c}^\infty dE\sigma_{\bf p}(E)e^{-ET}}
{\vert \psi_0 (0)\vert^2e^{-E_0T}\langle0|\hat{O_i}|0\rangle}. \nonumber
\end{eqnarray}
We shall use here the exact answers for the harmonic oscillator:
\begin{equation}
  E^{\rm har}_0=3/2; \ \ \  |\psi^{\rm har}_0(0)|^2=1/(2\pi)^{3/2};\ \ \
\langle0|\hat{O_1}|0\rangle_{\rm har}=1/3 E^{\rm har}_0; \ \ \ \
\langle0|\hat{O_2}|0\rangle_{\rm har}=1/2 E^{\rm har}_0.
\label{exhar}
\end{equation}

Let us consider the sum rules for the matrix elements
 $ \langle0|\hat{O_1}|0\rangle_{\rm har}$
and $\langle0|\hat{O_2}|0\rangle_{\rm har}$.
For the operator $\hat{O}_1$ both  functions $C_1^{\rm har}$ and 
$C_{\rm p1}^{\rm har}$are depicted for the energy
threshold $E_c=2$ (Fig.\ \ref{dualh1}). An important fact is immediately
noticeable. The true continuum is negative and cannot be approximated by
any positive asymptotics. Hence, no duality persists up to the first
exited state.
Note also that although the continuum contribution to the theoretical part of
the sum rule  is almost negligible (less than 5\%), the real
contribution of the exited states is significant and is about 40\% of the
ground state. Our choice of $E_c=2$ was motivated by standard guess --
it is a midpoint between two observed states $E_0=3/2$ and $(E_0+E_1)/2=5/2$.
The displayed picture is not sensitive to the threshold variation and the
duality is broken for any threshold parameter.

Let us illustrate how  duality breaking becomes
 fatal for the sum rule. The sum 
rule is obtained by transforming the continuum in equation (\ref{sumtri}) to
the l.h.s. and then dividing the expression by the two-point sum rule 
(eq. (\ref{specnew})):
\begin{equation}
  \label{tre}
\langle0|\hat{O_i}|0\rangle=\frac{S_i(T/2,T/2)-
\int_{E_c}^\infty dE\sigma_{\bf pi}(E)e^{-ET}}
{S_0(T) - \int_{E_c^0}^\infty dE\sigma_{\bf p0}(E)e^{-ET}}.
\end{equation}
 Here $S_i$ and $S_0$ are obtained by Perturbation Theory. Like in the 
previous case of two-point functions we have to determine a fiducial region.
The same conditions of the ground state dominance and the control over
the power expansions are applied. Of course, for the three-point sum rule
(\ref{tre}) we obtain two fiducial domains. One corresponds to the numenator
(three-point part) and another -- to the denominator (two-point part). The 
final window is then obtained in the matching region. In some three-point
sum rules discussed below continuum contributions are not small. Thus, the
standard prescription of the method (requirement for the continuum to be less
than 30\%) may lead to a situation when the window almost
shrinks to a point. In such cases we increase the bound up to 50\%.
An important notice is that three-point sum rules (\ref{tre}) depend on two
(in general independent) threshold parameters $E_c^0$ and $E_c$. While $E_c^0$
has to be fitted by the corresponding two-point sum rule, $E_c$ is a varying
parameter of the three-point sum rules. In practice, one usually takes both
thresholds equal. Below we present some arguments showing that in reality
$E_c$ is likely to be less than its two-point partner $E_c^0$. Like for the 
operator $\hat{O}_1$ (see above), in all three-point functions, 
which display the duality breaking, we take for the energy threshold $E_c$
the value, which is somewhat close to midpoint between two lowest observed 
states. In all these cases, the sum rules appear to be almost non-sensitive 
to the threshold variation and our main conclusions on duality violation
remain to be valid.

Fig.\ \ref{SL1}(a)  shows the sum rule for 
$ \langle0|\hat{O_1}|0\rangle_{\rm har}$ with the energy threshold
$E_c=2.5$. Within the window ($0.5\le T\le 1.6$) the answer given by the sum
rule is about 45\% off from the exact one (\ref{exhar}). Thus, the sum rule
leads to completely wrong prediction. In the Ref. \cite{BSU}, it was argued
that this failure is due to nondiagonal transitions, which are negative 
and numerically large. These transitions are not sufficiently suppressed
and they produce a strong influence on the sum rule. Sign alternating nature
of the exact spectral density $\sigma_h(E)$ leads to the duality breaking at
high energies.  To illustrate the point, we  include in the sum rule
explicitly a few low lying resonances.  Correspondingly, the continuum 
threshold $E_c$  rises:

\begin{equation}
  \label{sumN}
  \int_{E_c}^\infty dE\sigma_{\bf pi}(E)e^{-ET} \rightarrow 
\sum_{k=1}^N a_k^i e^{-\tilde{E}_kT}+\int_{E_c^N}^\infty 
dE\sigma_{\bf pi}(E)e^{-ET}.
\end{equation}
Here, $k$ runs over a number $N$ of the first low lying resonances in 
equation (\ref{kok}). The residues are denoted by $a_k^i$, while the 
resonances
are ordered by their energy levels $\tilde{E}_k$. 
For the case of the harmonic oscillator exact analytic expressions
for the energy levels $\tilde{E}_k$ and the residues of the interest $a_k^i$
are known \cite{BS}. The energy  $\tilde{E}_k=3/2+k$. 
\begin{eqnarray*}
  a_k^1&=&\frac{1}{(2\pi)^{3/2}}
\frac{4k+3}{6}\frac{1}{2^{2k}}\frac{(2k+1)!}{(k!)^2}, \ \ k-\rm{even}; \\
a_k^1&=&(-1)\frac{1}{(2\pi)^{3/2}}
\frac{2k+3}{3}\frac{1}{2^{2k}}\frac{(2k+1)!}{(k!)^2}, \ \ k-\rm{odd}.
\end{eqnarray*}
The sign alternating nature of the exact spectral function is clearly 
observed. 

We expect the energy threshold $E_c^N$ to be of order
$\tilde{E}_N$ - the energy of the last explicitly taken resonance. Sum rule
with continuum of the form (\ref{sumN}) is depicted for $N=3$ on 
Fig.\ \ref{SL1}(b) ($E_c^N=5$). The desired plateau is clearly restored.

The situation with the operator $\hat{O}_2$ strongly differs from the picture
described above. 
Despite the fact that the nondiagonal transitions are not vanishing, they 
 are of the same sign as the diagonal. Here we use the equation of 
motion to obtain residues  $a_k^2$:
\begin{eqnarray*}
a_k^2=\frac{\tilde{E}_k}{(2\pi)^{3/2}}-\frac{3}{2} a_k^1, \ \ k-\rm{even};
 \ \ a_k^2=(-1)\frac{3}{2} a_k^1, \ \ k-\rm{odd}.
\end{eqnarray*}
(The factor 3/2 in front of $a_k^1$ is due to definition of the operator 
$\hat{O}_1$.)
 
Thus, the exact spectral function is always positive. This fact will be 
shown to be crucial for the duality to hold.  
In order to check the duality, the  continuum functions $C_2^{\rm har}$ and
$C_{\rm p2}^{\rm har}$, eq. (\ref{cont}), are depicted (Fig.\ \ref{dualh2}) 
for the optimal energy threshold $E_c=2$. Both  functions match 
excellently and the duality is established. Consequently, the sum rule for 
VEV of the kinetic energy
operator can be investigated (Fig.\ \ref{SL2}).  An
important remark is in order. In the case at hand, continuum dominates in
the sum rule (it saturates more than 50\%) and the window is practically 
absent.  The obtained
sum rule displays strong sensitivity to the continuum threshold. Thus, 
such a behaviour of the sum rule is much like the one obtained in QCD 
\cite{BB}.   
Nevertheless, fitting the energy threshold, the exact result (\ref{exhar}) 
can be easily reproduced: $ \langle0|\hat{O_2}|0\rangle_{\rm har}=0.5$
at $E_c=2$. At this point we disagree with the conclusions of Ref. \cite{BSU}
on the sum rule failure. In this paper, the continuum threshold was taken
the same as for the two-point function $E_c=2.5$. However, in the three-point
function  at hand, $E=2.5$ is the energy level of the first nondiagonal 
transition state.
Consequently, for the sum rule the value of the energy threshold eventually 
has to be taken lower.    

\subsection{Linear oscillator}

With the same emphasise on duality, let us investigate the three-point 
functions for the linear potential. The operator $\hat{O}_3$ plays now
a role of the potential and it is the virial partner of the 
operator $\hat{O}_2$.
Since no exact propagator is known, asymptotic expansions of the three-point 
functions (eq. \ref{kok}) are obtained perturbatively. Details of this
computations are presented in Appendix B:
\begin{eqnarray}
  \label{s1l}
  S_1^L(T/2,T/2)&=&\frac{1}{32\pi^{3/2}}\frac{1}{T^{1/2}}
\left(1-\left(\sqrt{\pi}-\frac{4}{3\sqrt{\pi}}\right) T^{3/2} + 
\left(\frac{247}{120}-\frac{16\sqrt{2}}{15}\right) T^3 +\cdots\right);
\nonumber \\ \nonumber \\
 S_2^L(T/2,T/2)&=&\frac{3}{16\pi^{3/2}}\frac{1}{T^{5/2}}
\left(1-\frac{4}{3\sqrt{\pi}} T^{3/2} + 
\left(\frac{16 \sqrt{2}-19}{12}\right) T^3 +\cdots\right);  \\
\nonumber \\
S_3^L(T/2,T/2)&=&\frac{1}{4\pi^2}\frac{1}{T}
\left(1-\left(\sqrt{2\pi}-\frac{7\sqrt{\pi}}{8}\right) T^{3/2}+ 
\left(\frac{35}{48}-\frac{5\pi}{64}\right) T^3 +\cdots\right).
\nonumber
\end{eqnarray}
The corresponding parton spectral densities are:
\begin{eqnarray}
\sigma^L_{\rm p1}(E)&=&\frac{1}{32\pi^2}E^{-1/2};   \nonumber \\ \label{spec1}
\sigma^L_{\rm p2}(E)&=&\frac{3}{16\pi^{3/2}}\left(\frac{4}{3\sqrt{\pi}}
E^{3/2}-\frac{4}{3\sqrt{\pi}}\right); \\
\sigma^L_{\rm p3}(E)&=&\frac{1}{4\pi^2}. \nonumber
\end{eqnarray}

The general picture with the three-point sum rules for the linear oscillator 
is very similar to the one of the harmonic oscillator. 
We start from the operator $\hat{O}_3$ and check  the duality first. 
Fig.\ \ref{dual1} shows  the continuum functions
$C^L_3$ and $C^L_{\rm p3}$ (eq. \ref{cont}); $E_c=3.3$. Again, 
the true continuum
is mostly negative and cannot be represented by the asymptotics.  A best fit 
would be reached in the ``no continuum'' approximation. Consider now  the
sum rule  for the VEV of the operator. The window for the
linear oscillator is moved to the left: $0.3\le T\le 0.7$. Comparing to the 
exact numerical result $\langle 0|\hat{O}_3|0\rangle^L_{\rm exact}=1.559$, 
one $\langle 0|\hat{O}_3|0\rangle^L=1.1\pm 0.1$
yield by the sum rule (Fig.\ \ref{SLIN1}(a)) is 35\% smaller.  We  
account here for the situation, when the positive diagonal transitions almost
cancel the negative nondiagonal matrix elements. This results in the fact that
the large number of resonances must be explicitly taken in eq. (\ref{sumN})
 in order
for the sum rule to be saturated. Fig.\ \ref{SLIN1}(b) presents the sum rule
at $N=11$ ( The numerical values for the residues $a_k$ are given
in the tables \ref{t2} and \ref{t3}).
The energy threshold is $E_c=7$ that  lies between third and fourth
energy levels. 

Consider now the operator $\hat{O}_1$.  The sum rule displays the same
problem with duality as for the harmonic oscillator. No duality persists 
up to the first exited state. Fig.\ \ref{SLIN3} shows the sum rule for VEV
of the operator together with an improved continuum model  (\ref{sumN}).
The plateau is restored when $N=11$ transitions are taken explicitly. 

Let us turn now  to the three-point function with  
$\hat{O}_2$ operator inserted. 
The corresponding hadron spectral function  is positive. As it 
was argued above,  positive
spectral functions do not cause  duality breaking.   The case
at hand confirms this statement. Although the sum rule strongly depends on 
the continuum threshold parameter, it, nevertheless, 
yields the correct  value
(Fig.\ \ref{SLIN2}): $\langle 0|\hat{O}_2|0\rangle^L=0.81\pm 0.01$, compared
to the exact $\langle 0|\hat{O}_2|0\rangle^L_{\rm exact}=0.779$. 
The optimal energy threshold
parameter is $E_c=2.8$, which is significantly below the threshold parameter
corresponding to the two-point sum rule. 

As a common property of the three-point functions with sign changing
spectral functions, we see that a large amount of resonances must be taken
into account explicitly. In other words, no duality is valid in the low energy
area. This effect is due to the fact , which was already mentioned. 
The window moves to the left, where exited states are not sufficiently 
suppressed.

\subsection{Linear + Coulomb model}

Coulomb term added to the potential improves a bit the situation
slightly throwing out the window to the right. However, the general picture 
of the duality breaking still persists. We now present our results for the
three-point sum-rules in ``Linear + Coulomb'' potential. 
 The three-point functions (\ref{kok}) are obtained by perturbation (see
Appendix B):

\begin{eqnarray}
& &S_1^{\rm lc}(T/2,T/2)=
\frac{1}{32\pi^{3/2}}\frac{1}{T^{1/2}}\times \nonumber \\
& &\left(1+b_0\frac{4(\pi-1)}{3\sqrt{\pi}}T^{1/2} +
b_0^2 1.358 T -b_0 1.575 T^2 
-\left(\sqrt{\pi}-\frac{4}{3\sqrt{\pi}}\right) T^{3/2} + 
\left(\frac{247}{120}-\frac{16\sqrt{2}}{15}\right) T^3\right );
\nonumber \\ \nonumber \\
& & S_2^{\rm lc}(T/2,T/2)= \label{s1lc} \\
& &\frac{3}{16\pi^{3/2}}\frac{1}{T^{5/2}}
\left(1+b_0\frac{4+2\pi}{3\sqrt{\pi}}T^{1/2}+b_0^2 1.97 T -b_0 1.38 T^2
-\frac{4}{3\sqrt{\pi}} T^{3/2} + 
\left(\frac{16\sqrt{2}-19}{12}\right) T^3 \right); \nonumber \\
\nonumber \\
& &S_3^{\rm lc}(T/2,T/2)= \nonumber \\
& &\frac{1}{4\pi^2}\frac{1}{T}
\left(1+b_0 1.679 T^{1/2}+ b_0^2 1.477 T -b_0 1.538 T^2
-\left(\sqrt{2\pi}-\frac{7\sqrt{\pi}}{8}\right) T^{3/2}+ 
\left(\frac{35}{48}-\frac{5\pi}{64}\right) T^3 \right).
\nonumber
\end{eqnarray}
The corresponding spectral densities are
\begin{eqnarray}
\sigma^{\rm lc}_{\rm p1}(E)&=&\frac{1}{32\pi^2}E^{-1/2};   \nonumber \\ 
\sigma^{\rm lc}_{\rm p2}(E)&=&\frac{3}{16\pi^{3/2}}
\left(\frac{4}{3\sqrt{\pi}}E^{3/2}
+b_0\frac{4+2\pi}{3\sqrt{\pi}} E +b_0^2 2.22 E^{1/2}
-\frac{4}{3\sqrt{\pi}} -b_0 0.78 E^{-1/2} \right); \label{speclc}  \\ 
\sigma^{\rm lc}_{\rm p3}(E)&=&\frac{1}{4\pi^2}
(1+b_0 0.948 E^{-1/2}). \nonumber
\end{eqnarray}

The behaviour of the sum rule for the kinetic energy operator is similar
 to ones of the harmonic and linear oscillators. Due to the
positiveness of the exact spectral function no duality breaking is accounted
for. Fig.\ {\ref{SLC2} shows the sum rule at the optimal
energy threshold $E_c=2.4$. The exact
value  $\langle 0|\hat{O}_2|0\rangle_{\rm exact}^{\rm lc}=0.972$ and it 
matches well the one given by the sum rule:
 $\langle 0|\hat{O}_2|0\rangle^{\rm lc}=0.96\pm 0.02$. 

Let us now consider the operators $\hat{O}_1$ and $\hat{O}_3$.
Below we plot  two graphs representing our results on the three-point
functions with duality breaking. 
Fig.\ \ref{SLC3} shows the usual sum rule
and the sum rule with $N=5$ explicitly taken resonances ($E_c=5.5$) 
 for VEV of the operator $\hat{O}_3$. Note
that this operator is no more a virial partner of the operator  $\hat{O}_2$.
Taking five transitions explicitly we restore the plateau at the exact
level $\langle 0|\hat{O}_3|0\rangle_{\rm exact}^{\rm lc}=1.401$.
Fig.\ \ref{SLC1} closes our analysis. It describes the sum rule and the 
sum rule with $N=3$ explicitly taken resonances
for  VEV of the operator $\hat{O}_1$;  
$\langle 0|\hat{O}_1|0\rangle_{\rm exact}^{\rm lc}=0.399$.

\section{Conclusion}
 
Motivated by the recent  work on three-point sum rules in QCD,
especially  connected to  Heavy Quark Effective Theory, 
 we studied the two- and three-point sum rules in
three different nonrelativistic quantum mechanical models with  
confining potentials. We have seen that though in all cases the two-point
sum rules work perfectly well, the three-point sum rules may fail.
Their predictions for the matrix elements of the operators $\hat{O}_{1,3}$
may differ by $30-50\%$ from the corresponding  true values.
The reason of the 
failure is the breakdown of duality. The theoretical spectral function 
is always smooth and positive. On the other hand, we have seen by explicit 
calculation that the ``phenomenological'' spectral density is wildly 
oscillating and even has a sign changing component due to the sign changing 
nondiagonal transitions. We have seen that, though the diagonal 
transitions between radial excitations have positive transition constants,
the nondiagonal transitions always have negative sign. Consequently, 
duality does not work for the first several 
resonances, even if one  understands the duality in the a ``generalised''
sense (i.e., even after the integration in the direction orthogonal to the 
diagonal). The averaged ``hadron'' density 
strongly differs from the ``theoretical'' one. 
It seems that the duality starts to work for energies high enough. 
The corresponding threshold lies near the 3d-4th resonance and depends on
the model and on the operator. However, this is of no practical interest,
because one cannot separate the leading resonance contribution in order to
apply the sum rules. 
The standard continuum model does not work, and the smooth
theoretical spectral density strongly differs from  real one.
 For the two-point sum rules the whole situation is quite contrary: 
the duality does work in all known examples (section II).

We have also seen that the absence of the continuum has no relation
to the validity of the sum rule. In fact, small continuum contribution
may arise from the mutual cancellation of positive (diagonal) and 
negative (nondiagonal) transitions. Such a behaviour is displayed by
 the sum rules for the matrix elements of the operators $r$ and $r^2$.
 Moreover, it seems that the 
absence of the continuum contribution is a  general feature of the 
sum rules whose right hand side is contaminated by the sign changing 
transitions. 
On the other hand,  
sum rules for the matrix element of the
operator $-\partial^2$ work sufficiently well,
despite the fact that they are dominated by large continuum. The duality
holds thanks to the positiveness of the spectral density. This property
is very similar to that 
for the three-point sum rules for the transitions under the action of 
Hamiltonian, where the duality is also not violated \cite{BSU}.

One may conclude that the duality breaking is a general 
feature of the quantum mechanical analogue of the QCD sum rules in the
case of  sign changing nondiagonal transitions. Furthermore, our
study suggests that  the duality
still holds if all transitions are positive.
 However, the continuum  contribution to the sum rule is likely
to be large and the result may heavily depend on the continuum threshold.  

Our results were obtained for the nonrelativistic quantum mechanical 
models. It will be very interesting to check if our picture of the 
duality breaking still holds for the relativistic analogues of the 
models A, B, and C. 

Unfortunately, we  do not know yet, what may be implications of our results 
in real QCD.
Several conclusions, however, can be reached.
 First, one must be very careful
in presence of the sign alternating transitions contributing to the 
polarisation operators.
 Second,  the smallness of the
continuum contribution is not always of a good omen, and may occur 
due to complicated cancellations in the right hand side of the sum rule.
Nevertheless, the situation in QCD may be considerably better.
Borel transform in QCD may suppress nondiagonal transitions
 stronger than in quantum mechanics, and this may  
 lead to the smallness 
of the contamination. Perhaps, the degree of  contamination 
 depends
on the matrix element one calculates and the form of the sum rule chosen.

 Certainly,  further work is needed;
in particular, it is desirable  to investigate 
 whether the complete relativistic calculation 
improves the situation with duality. Detail examination of whether in the 
three-point QCD and HQET sum rules there are indeed sign changing 
transitions is required. The problem of their relative suppression must
be carefully studied.
 
Our work was essentially motivated by   big 
discrepancies  among values of the matrix element 
$\langle B\vert \vec{D}^2/(2m_B)\vert B\rangle$
obtained using different sum rules (see Refs. \cite{Neubert,BB}) 
Consequently, we work with  simple nonrelativistic   
analogues for the potential models of B-mesons and 
study the sum rules for  nonrelativistic analogues of the HQET
operator $\vec{D}^2/(2m_B)$
(and  operators related to $\vec{D}^2/(2m_B)$  by the virial theorem).
Our results imply 
that the value for the matrix element of the operator $\vec{D}^2/(2m_B)$
from  Ref. \cite{Neubert} is underestimated. 
On the other hand, the situation in quantum mechanics seems to be close
to  that considered in Ref.  \cite{BB}. However, the strong dependence
on the continuum threshold implies that the accuracy of the relevant sum rule
may be quite low. Moreover, in the text, we argued that the energy threshold
of the three-point sum rule should be taken lower (about 20\%, 
according to our
experience) than that of the
corresponding two-point sum rule. The reason is that the contribution 
of the first nondiagonal transition to the polarization operator is
suppressed by the factor $\exp [-(E_1+E_0)T/2]$, while the contributions
of the first diagonal transition in the three-point sum rule as well as
of the first excited state in the two-point sum rule  
are suppressed by  $\exp [-E_1 T]$.  
Consequently, we decided to reexamine
the sum rule of Ref. \cite{BB}. In the latter, 
the energy threshold for the 
three-point sum rule ($\omega_0\sim 1-1.2$ {\rm GeV})
 was taken exactly the same  
as for the two-point sum rule. We have seen above 
that this treatment of the sum
rule may lead to a significant overestimation of the matrix element. 
We investigated the leading-order 
sum rule (eqs. (3.10) and (3.12) of 
Ref. \cite{BB}) for the kinetic energy operator for  various three-point
thresholds  $\omega_1\sim 0.8$ {\rm GeV}. 
The results of our analysis are depicted 
on Fig.\ \ref{braun}.  The sum rule yields the following value for the matrix 
element of the kinetic energy operator:
$\langle B|\vec{D}^2/(2m_B)|B\rangle =-0.3\pm 0.1$ {\rm GeV$^2$} compared
to $\langle B|\vec{D}^2/(2m_B)|B\rangle =-0.6\pm 0.1$ {\rm GeV$^2$} for
$\omega_1\sim 1$ {\rm GeV}.
The obtained value is in a good 
agreement with the ones obtained in Ref. \cite{KLW,Chernyak}.   
(Of course, this result must be considered not as  a QCD sum rule prediction,
but rather as an indication that there is no contradiction between Refs. 
\cite{KLW,Chernyak} and QCD sum rule approach).

After this research was finished we learned that the quantum mechanical
duality was studied in Ref. \cite{MK}  for $S\rightarrow P$ transitions.
However, no violation of duality was found in that type of transitions.

\acknowledgements 
The authors are indebted to M. Shifman for useful 
discussions.  We are thankful to S. Grigoryan for pointing out several minor
numerical mistakes and misprints in the first version of the paper.
This work was supported by the Israel Science Foundation
under the contract 94805  and by the Technion fund for promotion of 
basic research under the contract 090875 and by the Harry Werksman research
fund.

\appendix
\section{}

In the first part of this Appendix we present some known information about 
solutions of the three-dimensional Schroedinger equation in the case of linear
potential. Three operators relevant to our study are defined. Their 
matrix elements are computed numerically and represented in tables.
Second part of Appendix is devoted to the same analysis for the 
linear + Coulomb potential.

\subsection{Linear potential.} 

We look for exact solutions of the three-dimensional Schroedinger equation:

\begin{equation}
[-\partial^2 + \alpha r]\psi_n^L= \tilde{E}_n^L \psi_n^L.  
\label{A1}
\end{equation}
In the dimensionless variables 
\begin{eqnarray*}
  E_n^L= \tilde{E}_n^L/\alpha^{2/3}; \ \ \ x=r \alpha^{1/3}.
\end{eqnarray*}
equation (\ref{A1}) takes form:
\begin{equation}
  \label{A2}
[-\partial^2 + x]\psi_n^L= E_n^L \psi_n^L.    
\end{equation}
Since the wave-function of the orbitally exited states are 
vanishing in the origin, we restrict our 
analysis to  S-states only. The S-states of the equation (\ref{A2}) are given
by the normalized Airy functions:
\begin{equation}
  \label{A3}
 \psi_n^L= {\rm const} \times {\rm Ai}(x-E_n)/x.
\end{equation}
Imposing condition of nonsingularity on the wave-function in the origin, we
obtain the discrete spectrum:
\begin{eqnarray*}
  {\rm Ai}(-E_n)=0.
\end{eqnarray*}
Table \ref{t1} presents ten first energy levels of the problem.
Taking into account the well-known fact (Ref. \cite{Math})
\begin{eqnarray*}
{\rm Ai}(\xi \rightarrow -\infty)\sim \sin\left(\frac{2}{3}\xi^{2/3}+\pi/4\right),
\end{eqnarray*}
\noindent
we can determine a large $n$ asymptotic behaviour of the energy levels:
\begin{equation}
  \label{A4}
 \frac{2}{3} (E_n)^{3/2}=n\pi, \ \ \ \ \  n\rightarrow \infty.
\end{equation}
An important and very special property of the Airy functions is
\begin{equation}
\label{A5}
  |\psi_n^L(0)|^2=\frac{1}{4\pi},
\end{equation}
\noindent
and does not depend on $n$. However,
\begin{eqnarray*}
  \psi_n^L(0)=(-1)^n \frac{1}{\sqrt{4\pi}}.
\end{eqnarray*}

Below, we present the exact numerical results for the 
matrix elements of the following operators:
\begin{eqnarray*}
\hat{O}_1=x^2/6; \ \ \ \ \hat{O}_2=-\partial^2; \ \ \ \ \hat{O}_3=x.  
\end{eqnarray*}
The matrix elements are defined:
\begin{equation}
M^{ij}_k\equiv\langle i|\hat{O}_k|j\rangle
\equiv\int d^3 x\psi_i^{L*}(x)\hat{O}_k(x)\psi_j^L(x)
\label{A6}
\end{equation}
Note that by the equation of motion
\begin{equation}
\label{A7}
  M_2^{ij}= E_i \delta^{ij} - M_3^{ij}
\end{equation}
The lowest quarters of the
matrices $M_1$ and $M_3$ age given by tables \ref{t2} and \ref{t3} 
respectively.

\subsection{Linear + Coulomb potential}

\begin{equation}
[-\partial^2 + V(r)]\psi_n^{\rm lc}= \tilde{E}_n^{\rm lc} \psi_n^{\rm lc}.  
\label{A8}
\end{equation}
The potential $V(r)$ is taken of the form:
\begin{equation}
  \label{A9}
  V(r)= \alpha r - \frac{4}{3}\frac{ \alpha _s(r)}{r}; \ \
\ \ \  \alpha _s(r)=\frac{2\pi}{9 \ln(\delta+\gamma/r)}.
\end{equation}
Here, $\alpha _s(r)$ is a running coupling constant and it reflects the 
asymptotic freedom of the strong interaction. In order to pass to 
dimensionless variables the following change of variables is performed:
\begin{eqnarray*}
x=(2m\alpha)^{1/3} r;\ \ \ \ \ \ E^{\rm lc}=
\tilde{E}_n^{\rm lc}(2m/\alpha^2)^{1/3}.  
\end{eqnarray*}
In the new variables equation (\ref{A8}) reads:
\begin{eqnarray}
[-\partial^2 +x- b(x)/x]\psi_n^{\rm lc}= E_n^{\rm lc} \psi_n^{\rm lc}; 
\nonumber  \\   \label{Apr}
\\  \nonumber
b(x)=\frac{8\pi (4m^2/\alpha)^{1/3}}{27\ln(\delta+\gamma(2m\alpha)^{1/3}/x)}. 
\end{eqnarray}

In Appendix B we present a perturbative determination of the propagator for
the given problem. Unfortunately, this procedure cannot be performed 
(at least
analytically) if the running logarithm is present. Since we do not wish to
introduce any uncertainty due to the logarithm factorisation,
 we fix the running
coupling constant $\alpha_s$ at some value. 
A special choice of the parameters 
is not important for our analysis, but to be concrete we choose some 
quasi-realistic ones \cite{AMT}: 
the string tension \  $\alpha=0.14$ {\rm GeV}$^2$; \ \
$\delta=2; \ \ \gamma=1.87$ {\rm GeV}$^{-1}$; 
\ for the constituent mass we take \ \ 
$m=0.35$ {\rm GeV}. When related problems are investigated within QCD the 
running
coupling constant $\alpha_s$ is usually of order 0.3. 
In our study $\alpha_s$ is set to
be 0.28, which corresponds to some fixed point in space, namely $x_0=0.088$.
In the text, a parameter $b_0$ is used and it is defined as 
$b_0\equiv b(x_0)$ (\ref{Apr}).

In order to solve equation (\ref{Apr}),
we consider  the Coulomb potential as a 
perturbation. Solution of the nonperturbed problem was described above.
The Hamilton  (\ref{Apr}) is diagonalized in the basis of the 
wave-functions (\ref{A3}):

\begin{equation}
 \psi^{\rm lc}_n(x)=\sum_{k=1}^{N_L}C^k_n\psi^L_k(x) 
\label{A11}
\end{equation}

In order to obtain the low energy spectrum it is sufficient to take into
account only a few levels (here $N_L$ denotes the number of levels).
 Table \ref{t4} presents eigenvalues computed 
by the numerical diagonalization.

Products of the wave-functions in the origin are

\begin{equation}
  \label{A12}
  C^{\rm lc}_{nm}\equiv \psi^{\rm lc}_n(0)\psi_m^{\rm lc*}(0)
=\frac{1}{4\pi}\sum_{k,k'}^{N_L} C_n^k C_m^{k'} (-1)^{k+k'}.
\end{equation}

Coefficients $C^{\rm lc}_{nm}$ appear to be slowly changing 
monotonic functions of the
number of levels $N_L$ 
taken into account in equations (\ref{A11}) and (\ref{A12}).
In order to extract these coefficients 
$N_L=60$ was taken and then the results 
were extrapolated. To illustrate the point we plot $C^{\rm lc}_{00}$ as a 
function of $N_L$ (Fig.\ \ref{C^LC}).  This function slowly approaches
(as a power law) to its limiting value $C^{\rm lc}_{00}=1.83$. 
This procedure may lead to some small numerical underestimation, which 
we estimate not to exceed a five percent level. Table \ref{t5} 
is devoted to the coefficients $C^{\rm lc}_{nm}$.

As for the linear potential, we present tables of numerically calculated
exact matrix elements (tables \ref{t6}, \ref{t7}, and \ref{t8}).

\section{}

In this Appendix, we present a perturbative derivation of propagators
and matrix elements of interest. The potential $V(r)=\alpha r - \beta/r$ is
considered as a perturbation to the free particle motion.  To study ``pure''
linear potential $\beta$ must be set to zero. 

The ``free'' propagator is given by
\begin{equation}
  \label{B1}
  G_0(r,T,r',t')=\frac{1}{[4\pi(T-t')]^{3/2}} 
\exp\left[-\frac{(\vec{r}-\vec{r'})^2}{4(T-t')}\right].
\end{equation}
Note that in our units $2m=1$. This factor can be always restored in final 
expressions. The exact propagator $G^{\rm ex }$ can be expressed in a formal
perturbation series:

\begin{eqnarray}
G^{\rm ex }(r,T,r',t')=G_0(r,T,r',t')-
\int d^3s \int_{t'}^T d\tau G_0(r,T,s,\tau) V(s) 
G_0(s,\tau,r',t')+  \nonumber \\ \label{B2}\\
\int d^3s \int_{t'}^T d\tau \int d^3{s'}\int_{t'}^{\tau}
d\tau' G_0(r,T,s,\tau)V(s) G_0(s,\tau,s',\tau')V(s') 
G_0(s',\tau',r',t') + \cdots . \nonumber
\end{eqnarray}
In the following analysis we retain only terms up to the second order in 
perturbation. All terms of the order $\alpha^3,\beta^3,\alpha\beta^2,
\alpha^2\beta$ and
higher will be systematically omitted. We denote:

\begin{eqnarray}
   \label{B4}
&G^{\rm ex }&=G_0+\alpha G_1^L+\alpha^2G_2^L+\beta G_1^C+\beta^2G_2^C+
\alpha\beta (G^{\rm LC}_2+G^{\rm CL}_2);
\nonumber \\ 
  &G_1^L&=\langle G_0|s|G_0\rangle; \ \ \ \ G_2^L=\langle G_1^L|s|G_0\rangle;
\ \ \ \  G_1^C=\langle G_0|1/s|G_0\rangle; \nonumber \\
&G_2^C&=\langle G_1^C|1/s|G_0\rangle; \ \ \ \
G_2^{\rm LC}=\langle G_1^L|1/s|G_0\rangle; \ \ \ \ 
G_2^{\rm CL}=\langle G_1^C|s|G_0\rangle.  
 \end{eqnarray}
Here, brackets denote the integration over 
$s$ and $t$ defined in eq. (\ref{B2}).
Unfortunately, these integrations cannot be done in closed form. However,
the time integration can be performed. This is done with the aid of the 
following integrals (Ref. \cite{FH}):
\begin{eqnarray*}
 \int_0^t d\tau \frac{1}{[(t-\tau )\tau ]^{3/2}}
\exp\left[-\frac{x^2}{(t-\tau )}-\frac{y^2}{\tau }\right]=
\frac{\sqrt{\pi}}{t^{3/2}}\frac{x+y}{xy}\exp\left[-\frac{(x+y)^2}{t}\right];
\\ \\
\int_0^t d\tau \frac{1}{(t-\tau )^{3/2}\tau^{1/2}}
\exp\left[-\frac{x^2}{(t-\tau )}-\frac{y^2}{\tau }\right]=
\frac{\sqrt{\pi}}{x\sqrt{t}}\exp\left[-\frac{(x+y)^2}{t}\right]. \ \ \ \ \ \
\end{eqnarray*}

After the time integration is removed, the angular integration can be easily 
performed as well. The remaining integrals over the radial components are
of a nonreducible form, and we cannot proceed any further. 
Finally we arrive at the following expressions for the propagator components:
\begin{eqnarray*}
G_1^L(r,T,0,t')&=&-\frac{1}{(4\pi)^{3/2}}\frac{r}{2(T-t')^{1/2}}
\left[ \int_1^\infty dx e^{-\frac{x^2 r^2}{4 (T-t')}} + 
e^{-\frac{r^2}{4 (T-t')}}\right]; \\ \\
G_2^L(r,T,0,t')&=&-\frac{1}{32\pi^{3/2}}\frac{1}{(T-t')^{1/2}}
\int_0^\infty  \frac{ds s^4}{r} \int_1^\infty dw \times \\ \\
& &\left[e^{-\frac{(r+s(1+w))^2}{4 (T-t')}} - 
e^{-\frac{(|r-s|+w s)^2}{4 (T-t')}} + \int_1^\infty dy y 
\left (e^{-\frac{(r+s(1+w y))^2}{4 (T-t')}}-
e^{-\frac{(|r-s|+w ys)^2}{4 (T-t')}}\right) \right]; \\ \\
G_1^C(r,T,0,t')&=& -\frac{1}{(4\pi)^{3/2}}\frac{1/r}{(T-t')^{1/2}}
\int_0^\infty  \frac{dx}{x}
\left[e^{-\frac{r^2(1+2x)^2}{4 (T-t')}}-
e^{-\frac{r^2(|1-x|+x)^2}{4 (T-t')}}\right]; \\ \\ 
G_2^C(r,T,0,t')&=&  
-\frac{1}{(4\pi)^{3/2}}\frac{1}{2(T-t')^{1/2}} \times \\ \\ & &
\int_0^\infty\frac{ds}{r}\int_0^\infty\frac{dx}{x}
 \int_{|1-x|+x}^{1+2x} dw \left[
e^{-\frac{(r+s(1+w))^2}{4 (T-t')}} - 
e^{-\frac{(|r-s|+w s)^2}{4 (T-t')}} \right]; \\ \\
G_2^{\rm LC}(r,T,0,t')&=& -\frac{1}{(4\pi)^{3/2}}
\frac{1}{4(T-t')^{1/2}}
\int_0^\infty\frac{ds s^2}{r}\int_1^\infty dw \\ \\
& &\left[\int_1^\infty dx x
\left( e^{-\frac{(r+s(1+w x))^2}{4 (T-t')}} 
- e^{-\frac{(|r-s|+w x s)^2}{4 (T-t')}}\right) +
e^{-\frac{(r+s(1+w))^2}{4 (T-t')}} - 
e^{-\frac{(|r-s|+w s)^2}{4 (T-t')}} \right]; \\ \\
G_2^{\rm CL}(r,T,0,t')&=& -\frac{1}{(4\pi)^{3/2}}
\frac{1}{2(T-t')^{1/2}}
\int_0^\infty\frac{ds s^2}{r}\int_1^\infty dw \int_0^\infty \frac{dx}{x} 
\times \\ \\
& &(1+2x)\left( e^{-\frac{(r+s(1+w (1+2x)))^2}{4 (T-t')}} 
- e^{-\frac{(|r-s|+ws(1+2x))^2}{4 (T-t')}}\right) \\ \\ & &
+ (|1-x|+x) \left( e^{-\frac{(r+s(1+w(|1-x|+x)))^2}{4 (T-t')}} - 
e^{-\frac{(|r-s|+w s (|1-x|+x))^2}{4 (T-t')}} \right).
\end{eqnarray*}
In order
to obtain equations (\ref{sl0}) and (\ref{slc0}), the limit $r\rightarrow 0$ 
is taken. At
this limit the propagator components (\ref{B4}) are completely calculated 
analytically and results of Ref. \cite{WDD} are recovered. 

Having in our disposal the exact propagator (\ref{B4}), we can proceed in 
computing the matrix element of the operators $\hat{O}_i(r)$. 
We are interested only in the following 
amplitudes: the free propagation from the point (0,0), in the 
point $(r,T/2)$, where the operator $\hat{O}_i(r)$ is inserted, 
and then there is the free motion to the point $(0,T)$.  

\begin{eqnarray}
\langle G^{\rm ex }(0,&T&,r,T/2)|\hat{O}_i(r)
|G^{\rm ex }(r,T/2,0,0)\rangle=\nonumber \\
   \langle G_0&|\hat{O}_i|&G_0\rangle +
2\alpha \langle G_1^L|\hat{O}_i|G_0\rangle 
+ \alpha^2\langle G_1^L|\hat{O}_i|G_1^L\rangle + 
2\alpha^2 \langle G_2^L|\hat{O}_i|G_0\rangle + 
2\beta \langle G_1^C|\hat{O}_i|G_0\rangle + \nonumber \\
&\beta^2& \langle G_1^C|\hat{O}_i|G_1^C\rangle  +  
2\beta^2 \langle G_2^C|\hat{O}_i|G_0\rangle +
2\alpha\beta \langle G_1^C|\hat{O}_i|G_1^L\rangle+
2\alpha\beta \langle G_2^{\rm LC}+G_2^{\rm CL}|\hat{O}_i|G_0\rangle. 
\label{B5}
\end{eqnarray}
In the equation (\ref{B5}) only three-dimensional $r$ integration is denoted 
by the brackets. The operator $\hat{O_i}(r)$ is one of the following 
operators:
\begin{eqnarray*}
  \hat{O}_1(r)=r^2/6; \  \hat{O}_2(r)=-\partial^2=\frac{1}{r^2}
\frac{\partial}{\partial r}\left(r^2\frac{\partial}{\partial r}\right);
\  \hat{O}_3(r)=r.
\end{eqnarray*}
For any matrix element in (\ref{B5}) $r$ enters as a polynomial times 
Gaussian. Thus, the $r$ integration can be  easily done. 
All remaining integrals have a fractional form and can be 
computed analytically or numerically.
Final results of these calculations are formulated in the text 
( \ref{s1l}, \ref{s1lc}).

\begin{table}
\begin{tabular}{||l|c|c|c|c|c|c|c|c|c||}  
\ $E_0^L$&$E_1^L$&$E_2^2$&
$E_3^L$&$E_4^L$&$E_5^L$&$E_6^L$&$E_7^L$&$E_8^L$&$E_9^L$  \\
\hline 
2.338    & 4.088 & 5.521 & 6.787 & 7.944 & 9.023 & 10.040& 11.009& 11.936&
12.829  \\ 
\end{tabular}        
\vspace{0.5 cm}
\caption{}
\label{t1}
\end{table}

\begin{table}
\begin{tabular}{||l||c|c|c|c|c||}  
$\langle i|\hat{O}_1|j\rangle_L$&  0  &  1  &  2  &  3  &  4  \\
\hline \hline
\ \ \ 0  &0.486&0.427 &-0.039& 0.010& -0.004       \\  
\hline
\ \ \ 1  &0.427&1.485 &0.950 &-0.075 & 0.018         \\
\hline
\ \ \ 2  & -0.039 & 0.950 &  2.709 & 1.556 & -0.116       \\
\hline
\ \ \ 3  & 0.010 & -0.075 & 1.556  & 4.094 & 2.223     \\
\hline
\ \ \ 4  & -0.004 & 0.018 & -0.116 & 2.223 &  5.610          \\ 
\end{tabular}        
\vspace{0.5 cm}
\caption{}
\label{t2}
\end{table}

\begin{table}
\begin{tabular}{||l||c|c|c|c|c||} 
$\langle i|\hat{O}_3|j\rangle_L$&  0  &   1    &    2    &    3    &   4 \\
\hline \hline
\ \ \ 0 & 1.559 & 0.653 & -0.197 & 0.101 & -0.064         \\  
\hline
\ \ \ 1 & 0.653 & 2.725 & 0.974 & -0.275 & 0.134        \\
\hline
\ \ \ 2 & -0.197 & 0.974 & 3.680 & 1.248 & -0.341       \\
\hline
\ \ \ 3 & 0.101 & -0.275 & 1.248 & 4.524 & 1.493           \\
\hline
\ \ \ 4 & -0.064 & 0.134 & -0.341 & 1.493 & 5.296              \\
\end{tabular}        
\vspace{0.5 cm}
\caption{}
\label{t3}
\end{table}

\begin{table}
\begin{tabular}{||l|c|c|c|c|c|c|c|c|c||} 
\ $E_0^{\rm lc}$ & $E_1^{\rm lc}$ & $E_2^{\rm lc}$ & 
$E_3^{\rm lc}$ & $E_4^{\rm lc}$ & $E_5^{\rm lc}$
 & $E_6^{\rm lc}$ & $E_7^{\rm lc}$ & $E_8^{\rm lc}$ & $E_9^{\rm lc}$  \\
\hline 
1.828  & 3.745 & 5.245 & 6.551 & 7.735 & 8.833 & 9.865 & 10.845& 11.783&
12.684  \\
\end{tabular}        
\vspace{0.5 cm}
\caption{}
\label{t4}
\end{table}

\begin{table}
\begin{tabular}{||l||c|c|c|c|c||} 
$C^{\rm lc}_{nm}$&  0  &   1    &    2    &    3    &   4 \\
\hline \hline
\ \ \ 0 &1.83  & -1.66 & -1.58 & 1.53  & 1.47         \\  
\hline
\ \ \ 1 &-1.66  &1.50  & 1.44 &  -1.39 & -1.35       \\
\hline
\ \ \ 2 &-1.58  &1.44  & 1.37 &  -1.34 & -1.31       \\
\hline
\ \ \ 3 & 1.53  &  -1.39 &  -1.34 & 1.32 &  1.28          \\
\hline
\ \ \ 4 &1.47  &-1.35  & -1.31 &1.28  &  1.25           \\
\end{tabular}
\vspace{0.5 cm}
\caption{}
\label{t5}
\end{table}        

\begin{table}
\begin{tabular}{||l||c|c|c|c|c||} 
$\langle i|\hat{O}_1|j\rangle_{\rm lc}$&  0  &   1    &    2    &    3    &  
 4 \\
\hline \hline
\ \ \ 0 &0.399  & 0.367 &0.042  & -0.013 & -0.006         \\  
\hline
\ \ \ 1 & 0.367 &1.363  &-0.874  & 0.079 &  0.021      \\
\hline
\ \ \ 2 & 0.042 & -0.874 & 2.569  & 1.471 &   0.120     \\
\hline
\ \ \ 3 & -0.013 & 0.079 & 1.471 & 3.942 &   -2.136          \\
\hline
\ \ \ 4 & -0.006 & 0.021 & 0.120 &  -2.136 &  5.448           \\
\end{tabular}
\vspace{0.5 cm}
\caption{}
\label{t6}        
\end{table}

\begin{table}
\begin{tabular}{||l||c|c|c|c|c||} 
$\langle i|\hat{O}_2|j\rangle_{\rm lc}$&  0  &   1    &    2    &    3    &   4\\
\hline \hline
\ \ \ 0 &0.972  & -0.810 &-0.338  & 0.216 & 0.162         \\  
\hline
\ \ \ 1 &-0.810  &1.483  &1.093  &-0.390  & -0.237       \\
\hline
\ \ \ 2 &-0.338  &1.093  &1.934   &-1.348  &-0.445        \\
\hline
\ \ \ 3 &0.216  & -0.390 & -1.348 &2.342  & 1.582           \\
\hline
\ \ \ 4 & 0.162 & -0.237 & -0.445 &  1.582 &  2.718           \\
\end{tabular}        
\vspace{0.5 cm}
\caption{}
\label{t7}
\end{table}

\begin{table}
\begin{tabular}{||l||c|c|c|c|c||} 
$\langle i|\hat{O}_3|j\rangle_{\rm lc}$&  0  &   1    &    2    &    3    &   4 \\
\hline \hline
\ \ \ 0 &1.401  &0.607  & 0.197 & -0.105 & -0.068         \\  
\hline
\ \ \ 1 & 0.607 & 2.615 &-0.929  &0.271  & 0.136       \\
\hline
\ \ \ 2 & 0.197 & -0.929 & 3.591 &1.206  & 0.336       \\
\hline
\ \ \ 3 & -0.105 & 0.271 &1.206  & 4.447 & -1.454           \\
\hline
\ \ \ 4 & -0.068 & 0.136 & 0.336 &  -1.454&  5.227           \\
\end{tabular}        
\vspace{0.5 cm}
\caption{}
\label{t8}
\end{table}
\begin{figure}[htbp]
\epsfxsize=0.4\textwidth
\epsfysize=0.4\textwidth
\centerline{\epsffile{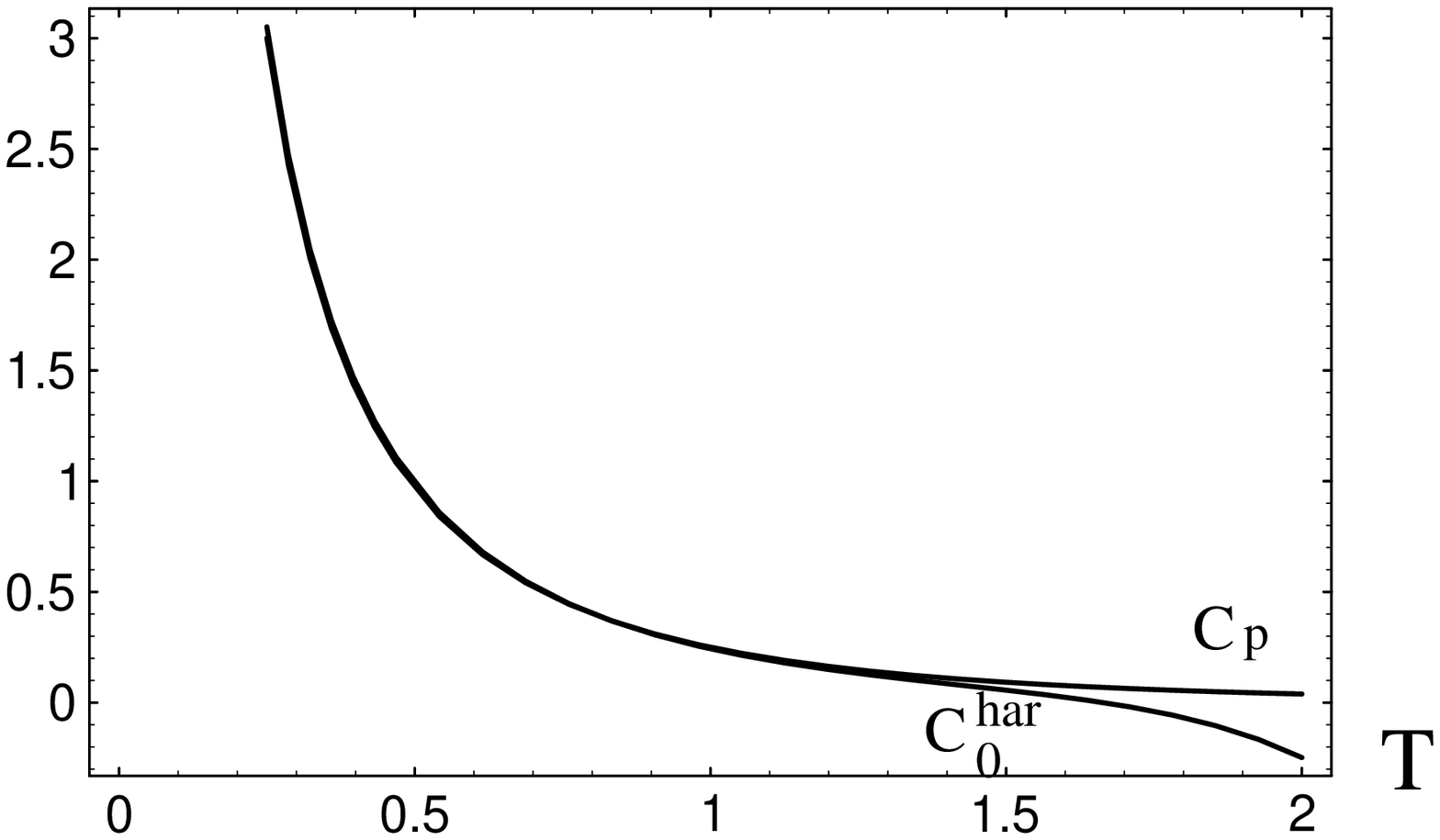}}
\vspace*{-1.5cm}
\caption{Harmonic oscillator. Duality for the two-point function. Continuum
functions $C^{\rm har}_0$ and $C_{\bf p}$ are plotted vs. $  T$. 
The energy threshold $E_c=2.6$.}
\label{S0}
\end{figure}
\begin{figure}[htbp]
\epsfxsize=0.4\textwidth
\epsfysize=0.4\textwidth
\centerline{\epsffile{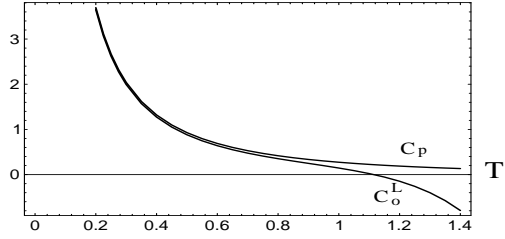}}
\vspace*{-1.5cm}
\caption{Linear oscillator. Duality for the two-point function. Continuum
functions $C^L_0$ and $C_{\bf p}$ are plotted vs. $T$. 
The energy threshold $E_c=3.4$.}
\label{SL0}
\end{figure}
\begin{figure}[htbp]
\epsfxsize=0.4\textwidth
\epsfysize=0.4\textwidth 
\centerline{\epsffile{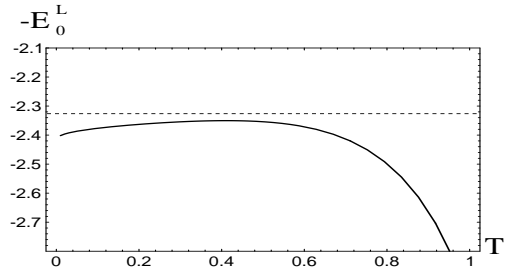}}
\vspace*{-1.5cm}
\caption{Linear oscillator. Sum rule for the ground state energy. 
 The dashed line corresponds to the exact value
$E_{0\rm exact}^L=2.338$. The energy threshold $E_c=3.4$.}
\label{fig2a}
\end{figure}
\begin{figure}[htbp]
\epsfxsize=0.4\textwidth
\epsfysize=0.4\textwidth
\centerline{\epsffile{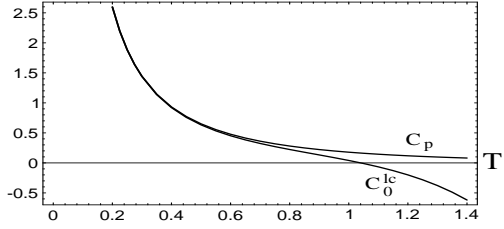}}
\vspace*{-1.5cm}
\caption{Linear + Coulomb potential. Duality for the two-point function.
Continuum functions $C^{\rm lc}_0$ and $C^{\rm lc}_{\bf p}$ are plotted vs. 
$T$. The energy threshold $E_c=2.9$.}
\label{SLC0}
\end{figure}
\begin{figure}[htbp]
\epsfxsize=0.4\textwidth
\epsfysize=0.4\textwidth
\centerline{\epsffile{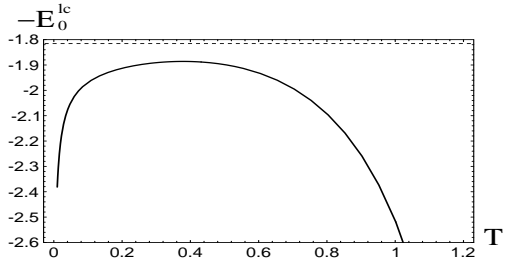}}
\vspace*{-1.5cm}
\caption{Linear + Coulomb potential. Sum rule for the ground state energy.
 The dashed line corresponds to the exact value
$E_{0\rm exact}^{\rm lc}=1.828$. The energy threshold $E_c=2.9$.}
\label{fig3a}
\end{figure}
\begin{figure}[htbp]
\epsfxsize=0.4\textwidth
\epsfysize=0.4\textwidth
\centerline{\epsffile{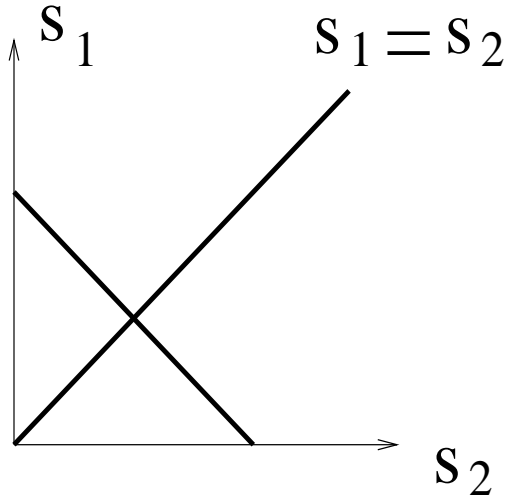}}
\vspace*{-0cm}
\caption{$(s_1, s_2)$ plane. Direction orthogonal to the diagonal is a
direction of the integration for the generalized duality.}
\label{spec}
\end{figure}
\begin{figure}[htbp]
\epsfxsize=0.4\textwidth
\epsfysize=0.4\textwidth
\centerline{\epsffile{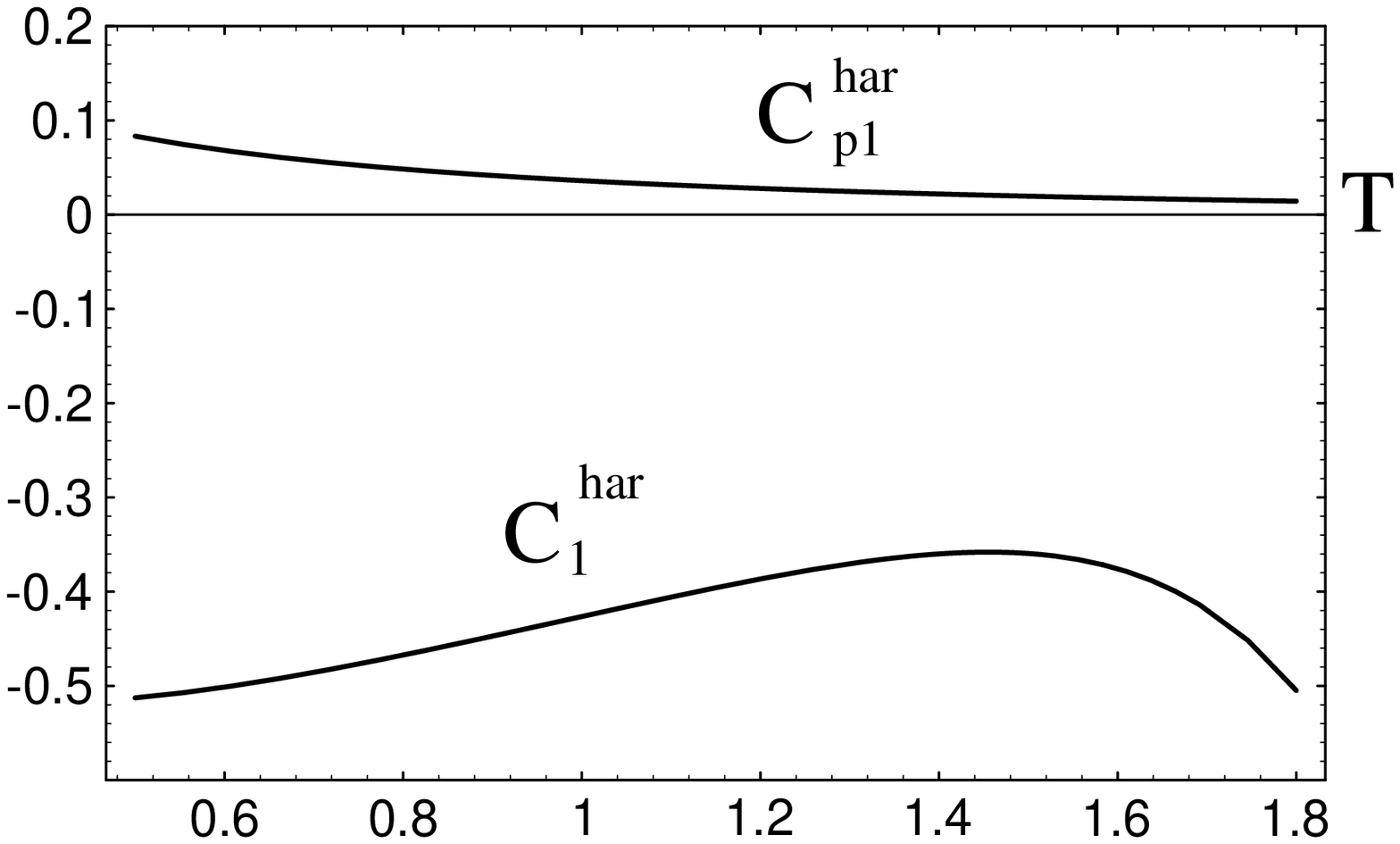}}
\vspace*{-1.5cm}
\caption{Harmonic oscillator. Duality for the three-point function. Insertion
of the operator $\hat{O}_1$. Continuum functions $C^{\rm har}_1$ and 
$C^{\rm har}_{\rm p1}$  are plotted vs. $T$. The energy threshold $E_c=2$.}
\label{dualh1}
\end{figure}
\begin{figure}[htbp]
\epsfxsize=0.4\textwidth
\epsfysize=0.4\textwidth
\centerline{\epsffile{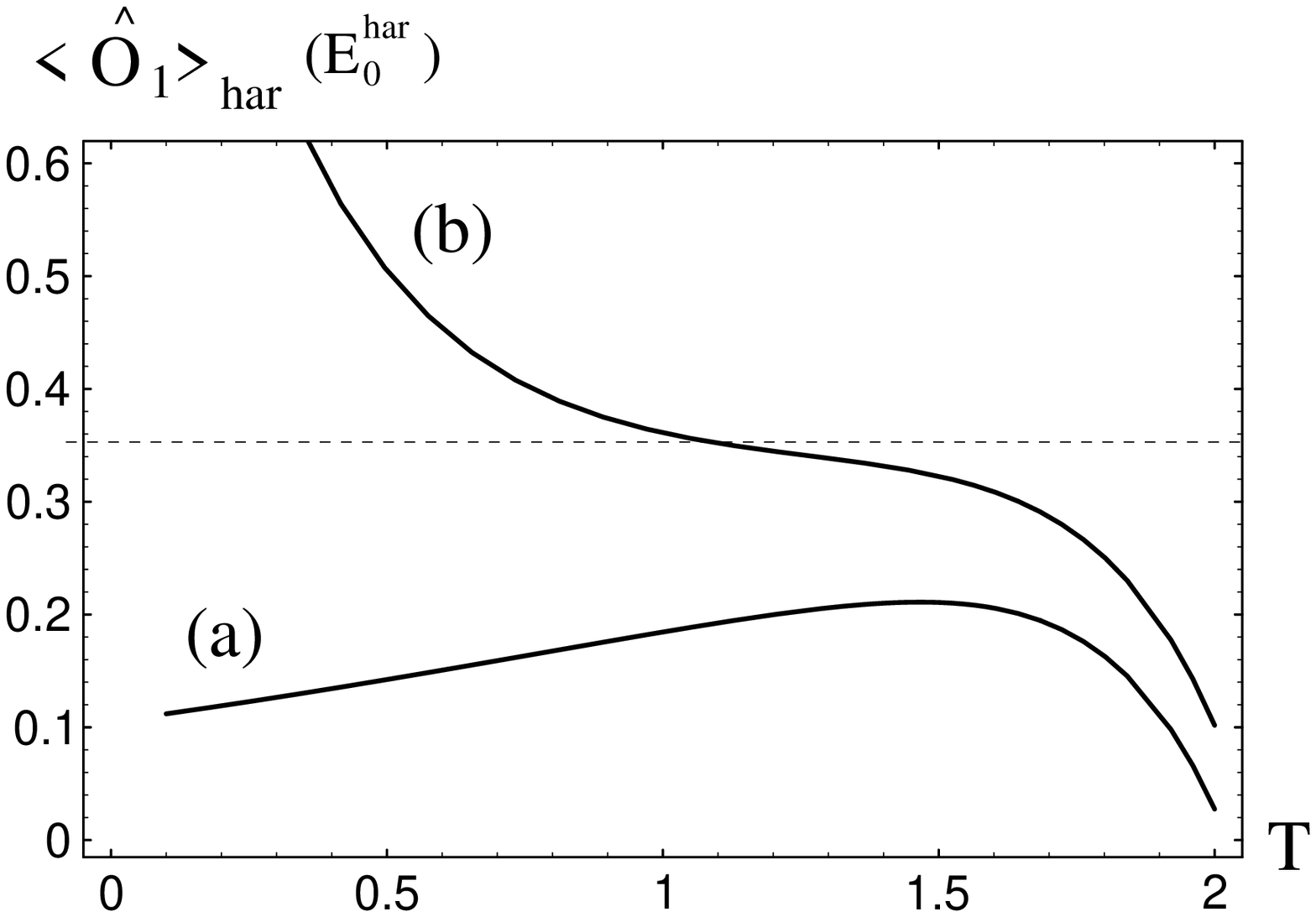}}
\vspace*{-1.5cm}
\caption{ Harmonic oscillator. (a) --
three-point sum rules for VEV of the operator
$\hat{O}_1$. The energy threshold $E_c=2$. (b) --
three-point sum rules for VEV of the operator
$\hat{O}_1$ with N=3 explicitly taken resonances.
The energy threshold $E_c^N=5$. The dashed line corresponds to the exact value
$\langle 0|\hat{O}_1|0\rangle_{\rm exact}^{\rm har}=1/3 
E_0^{\rm har}$.}
\label{SL1}
\end{figure}
\begin{figure}[htbp]
\epsfxsize=0.4\textwidth
\epsfysize=0.4\textwidth
\centerline{\epsffile{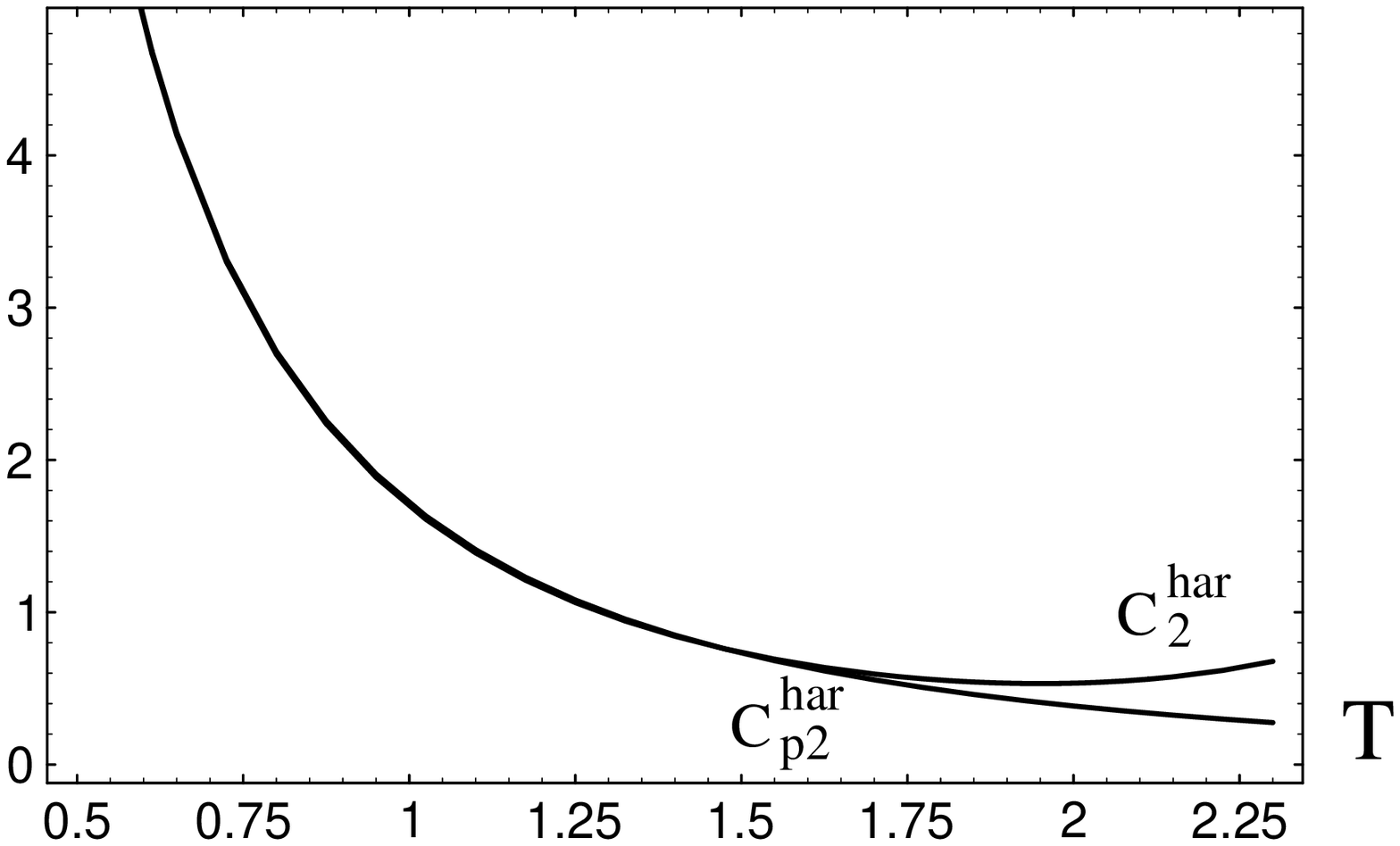}}
\vspace*{-1.5cm}
\caption{Harmonic oscillator. Duality for the three-point function. Insertion
of the operator $\hat{O}_2$. Continuum functions $C^{\rm har}_2$ and 
$C^{\rm har}_{\rm p2}$  are plotted vs. $T$. The energy threshold $E_c=2$.}
\label{dualh2}
\end{figure}
\begin{figure}[htbp]
\epsfxsize=0.4\textwidth
\epsfysize=0.4\textwidth
\centerline{\epsffile{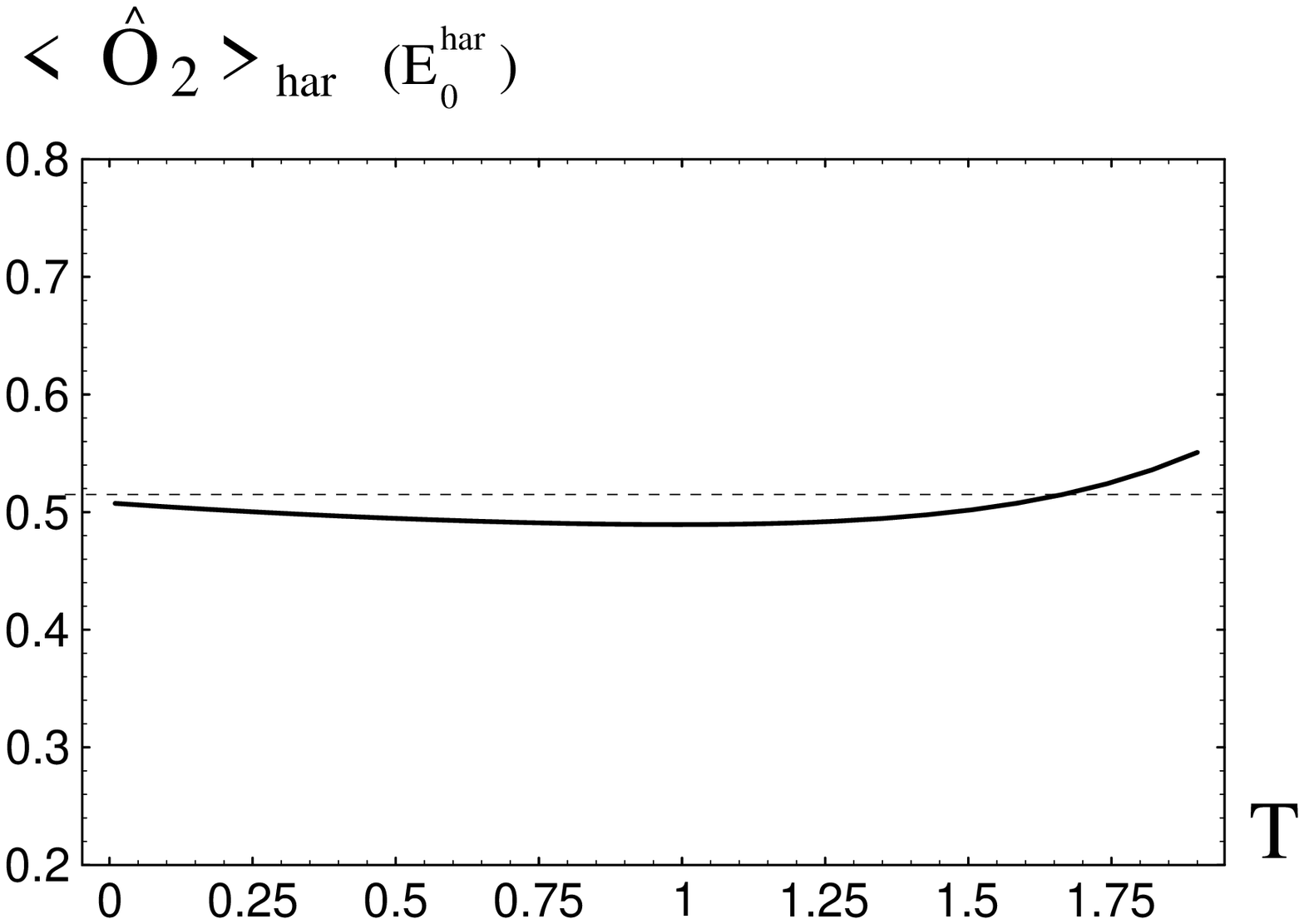}}
\vspace*{-1.5cm}
\caption{Harmonic oscillator. Three-point sum rules for VEV of the operator
$\hat{O}_2$.  The dashed line corresponds to the exact value
$\langle 0|\hat{O}_2|0\rangle_{\rm exact}^{\rm har}=1/2 
E_0^{\rm har}.$
The energy threshold $E_c=2.$.}
\label{SL2}
\end{figure}
\begin{figure}[htbp]
\epsfxsize=0.4\textwidth
\epsfysize=0.4\textwidth
\centerline{\epsffile{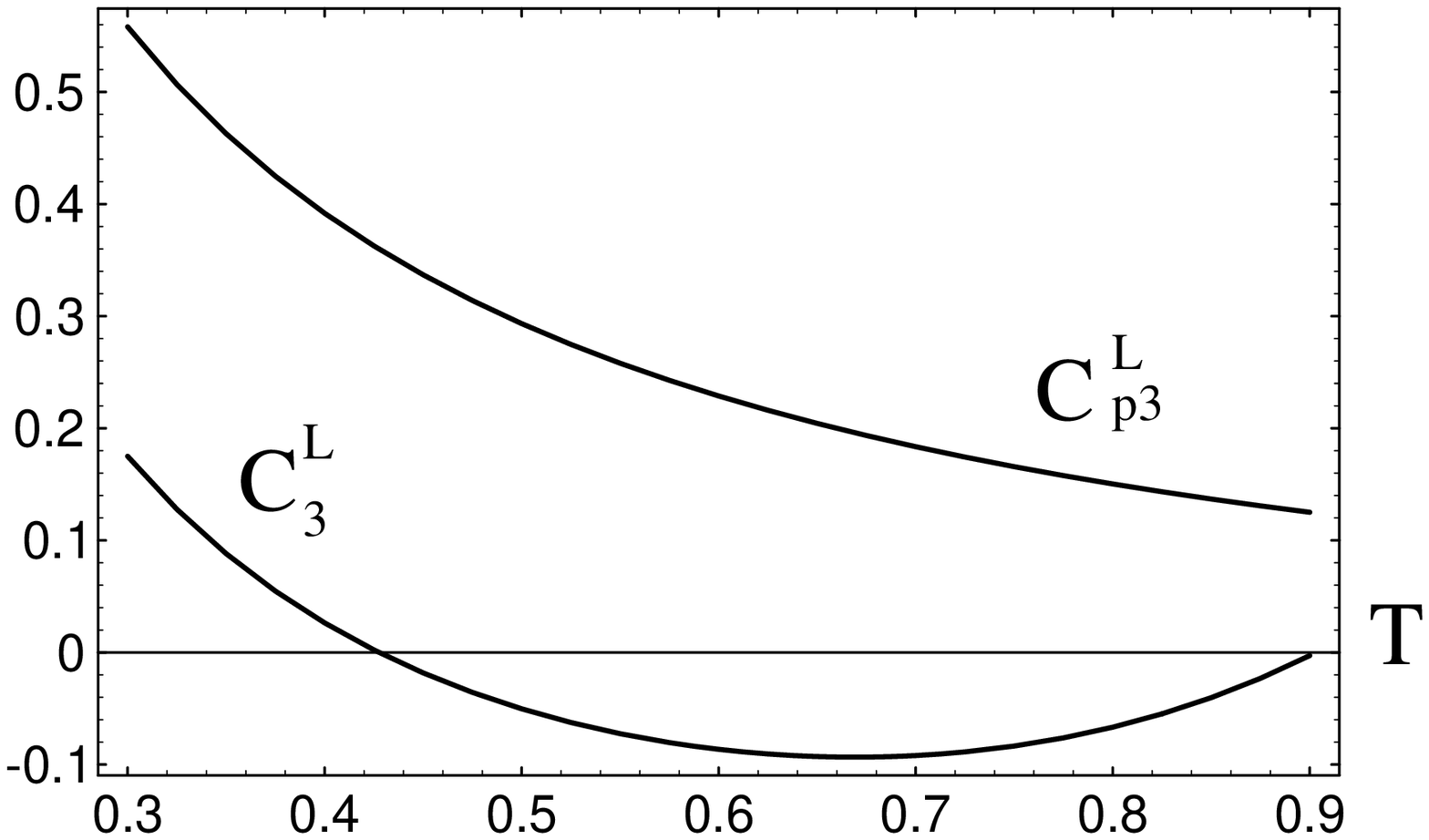}}
\vspace*{-1.5cm}
\caption{Linear oscillator. Duality for the three-point function. Insertion
of the operator $\hat{O}_3$. Continuum functions $C^{L}_3$ and 
$C^{L}_{\rm p3}$  are plotted vs. $T$. The energy threshold $E_c=3.3$.}
\label{dual1}
\end{figure}
\begin{figure}[htbp]
\epsfxsize=0.4\textwidth
\epsfysize=0.4\textwidth
\centerline{\epsffile{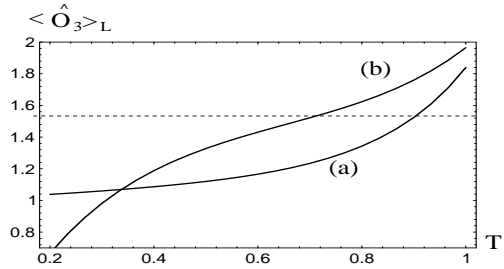}}
\vspace*{-1.5cm}
\caption{ Linear oscillator. 
(a) -- three-point sum rules for VEV of the operator
$\hat{O}_3$. The energy threshold $E_c=3.3$.
(b) -- three-point sum rules for VEV of the operator
$\hat{O}_3$ with N=11 explicitly taken resonances. The energy threshold 
$E_c^N=7$.
The dashed line corresponds to the exact value 
$\langle 0|\hat{O}_3|0\rangle_{\rm exact}^L=1.559$.}
\label{SLIN1}
\end{figure}
\begin{figure}[htbp]
\epsfxsize=0.4\textwidth
\epsfysize=0.4\textwidth
\centerline{\epsffile{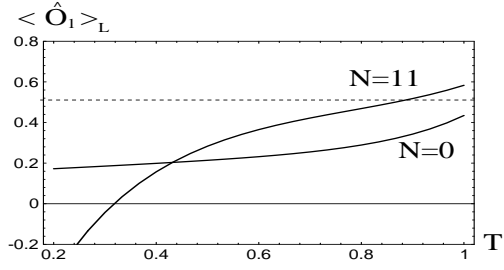}}
\vspace*{-1.5cm}
\caption{Linear oscillator. Three-point sum rules for VEV of the operator
$\hat{O}_1$ and the sum rule with  N=11 explicitly taken resonances.
 The dashed line corresponds to the exact value
$\langle 0|\hat{O}_1|0\rangle_{\rm exact}^L=0.486$. The energy thresholds 
$E_c=3.3$, $E_c^N=7$.  }
\label{SLIN3}
\end{figure}
\begin{figure}[htbp]
\epsfxsize=0.4\textwidth
\epsfysize=0.4\textwidth
\centerline{\epsffile{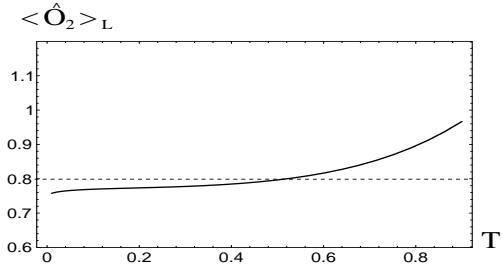}}
\vspace*{-1.5cm}
\caption{Linear oscillator. Three-point sum rules for VEV of the operator
$\hat{O}_2$.  The dashed line corresponds to the exact value
$\langle 0|\hat{O}_2|0\rangle_{\rm exact}^L=0.779$. The energy threshold 
$E_c=2.8$. }
\label{SLIN2}
\end{figure}
\begin{figure}[htbp]
\epsfxsize=0.4\textwidth
\epsfysize=0.4\textwidth
\centerline{\epsffile{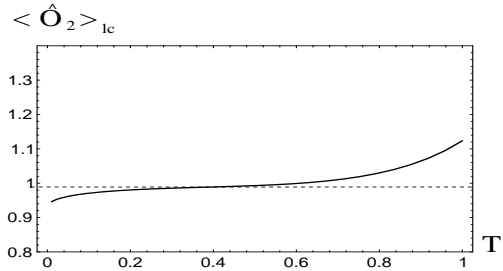}}
\vspace*{-1.5cm}
\caption{Linear + Coulomb model. 
Three-point sum rules for VEV of the operator
$\hat{O}_2$.  The dashed line corresponds to the exact value
$\langle 0|\hat{O}_2|0\rangle_{\rm exact}^L=0.972$. The energy threshold
$E_c=2.4$. }
\label{SLC2}
\end{figure}
\begin{figure}[htbp]
\epsfxsize=0.4\textwidth
\epsfysize=0.4\textwidth
\centerline{\epsffile{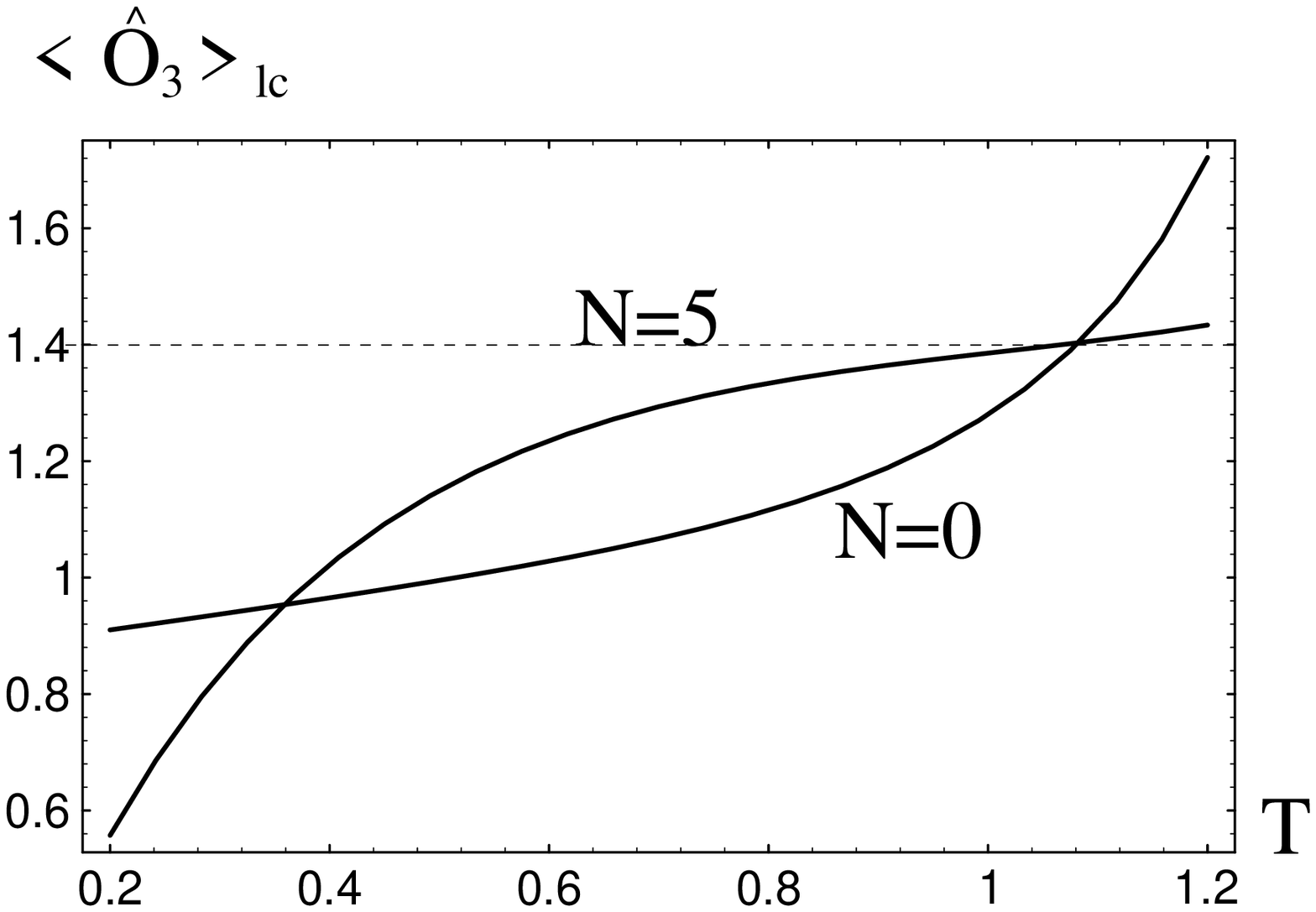}}
\vspace*{-1.5cm}
\caption{Linear + Coulomb model. 
Three-point sum rules for VEV of the operator
$\hat{O}_3$ and the sum rule with  N=5 explicitly taken resonances.
 The dashed line corresponds to the exact value
$\langle 0|\hat{O}_3|0\rangle_{\rm exact}^L=1.401$. The energy thresholds
$E_c=2.9$, $E_c^N=5.5$. }
\label{SLC3}
\end{figure}
\begin{figure}[htbp]
\epsfxsize=0.4\textwidth
\epsfysize=0.4\textwidth
\centerline{\epsffile{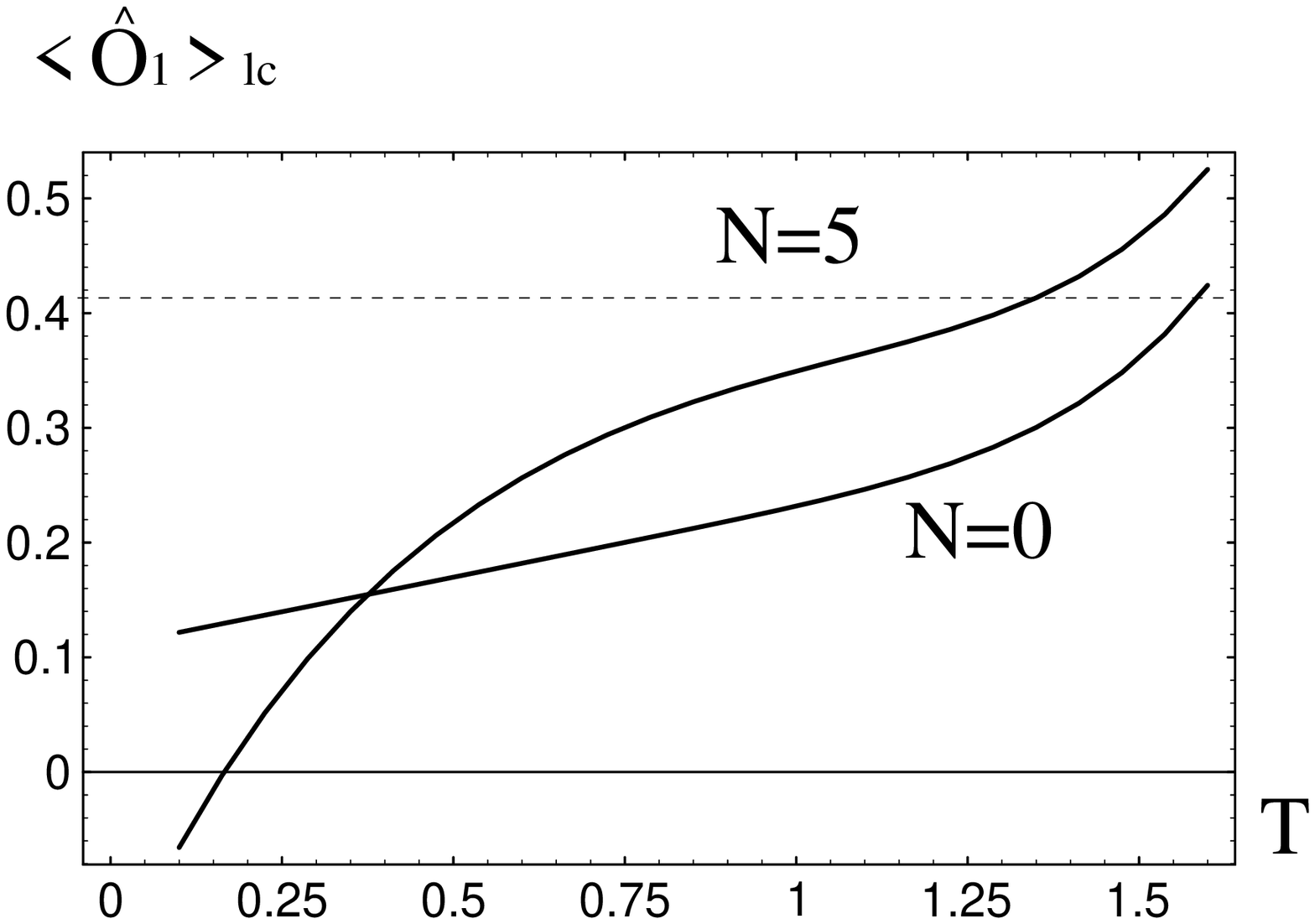}}
\vspace*{-1.5cm}
\caption{Linear + Coulomb model. 
Three-point sum rules for VEV of the operator
$\hat{O}_1$ and the sum rule with  N=5 explicitly taken resonances.
 The dashed line corresponds to the exact value
$\langle 0|\hat{O}_1|0\rangle_{\rm exact}^L=0.399$. The energy thresholds
 $E_c=2.9$, $E_c^N=5.5$. }
\label{SLC1}
\end{figure}
\begin{figure}[htbp]
\epsfxsize=0.5\textwidth
\epsfysize=0.5\textwidth
\centerline{\epsffile{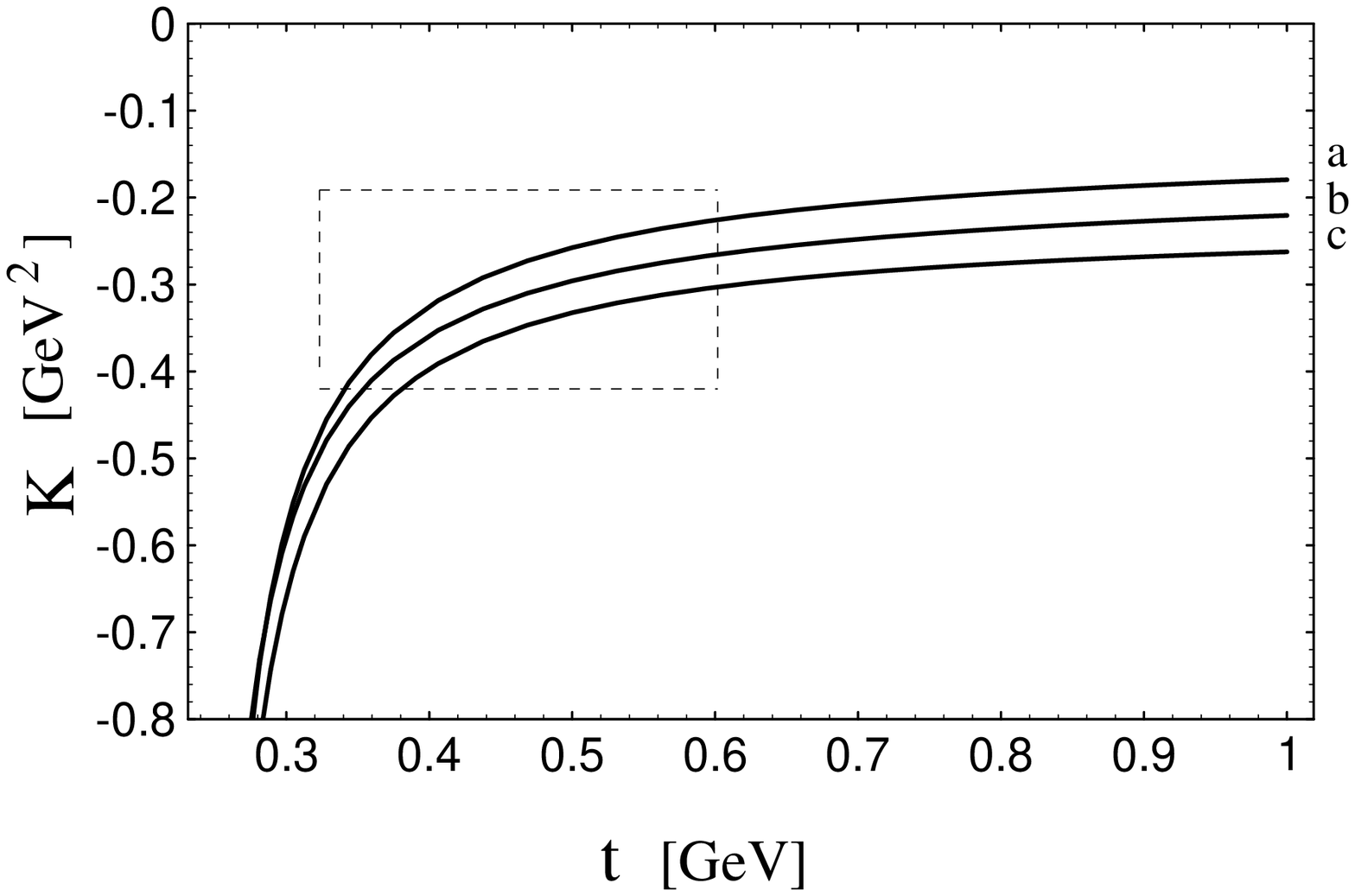}}
\vspace*{-1.5cm}
\caption{The Ball and Braun sum rule to leading-order as a function
of the Borel parameter $t$ for different values of the continuum thresholds:
a)  $\omega_0=1$ {\rm GeV}, $\omega_1=0.7$ {\rm GeV}; b)
 $\omega_0=1.2$ {\rm GeV}, 
$\omega_1=0.85$ {\rm GeV} ; c)  $\omega_0=1$ {\rm GeV},
 $\omega_1=0.8$ {\rm GeV}. 
The dashed line indicates the working region.}
\label{braun}
\end{figure}
\begin{figure}[htbp]
\epsfxsize=0.4\textwidth
\epsfysize=0.4\textwidth
\centerline{\epsffile{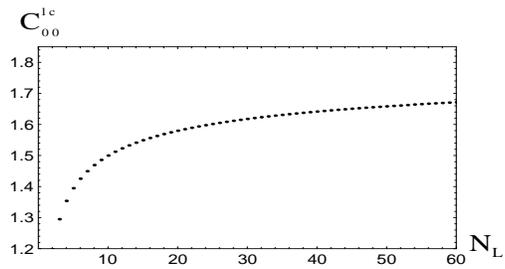}}
\vspace*{-1.5cm}
\caption{$C_{00}^{\rm lc}$ as a function of $N_L$.}
\label{C^LC}
\end{figure}
\end{document}